\documentclass[preprint, pra]{revtex4-1}
\usepackage{amsmath}
\usepackage{amssymb}
\usepackage{graphicx}
\usepackage{enumitem}

\newcommand{\x}{\vec{x}}
\newcommand{\GD}{G_{\Delta}( \x_1, \x_2 ) }
\newcommand{\GDA}{G_{\Delta}^{\text{A}}( \x_1, \x_2 )}

\renewcommand{\H}[1]{H(\x, \tau_#1;\x_#1 )}

\usepackage{bm}
\renewcommand{\vec}[1]{\boldsymbol{\mathbf{#1}}}
\usepackage[font={footnotesize}]{caption}
\usepackage{hyperref}
\hypersetup{pdftex,colorlinks=true,allcolors=blue}
\usepackage{hypcap}
\begin{document}

\title{Entanglement Entropy of Local Operators in Quantum Lifshitz Theory}
 
\author{Tianci Zhou}
\email{tzhou13@illinois.edu}
\affiliation{University of Illinois, Department of Physics, 1110 W. Green St. Urbana, IL 61801 USA}
 
\date{\today}

\begin{abstract}
We study the growth of entanglement entropy(EE) of local operator excitation in the quantum Lifshitz model which has dynamic exponent $z = 2$. Specifically, we act a local vertex operator on the groundstate at a distance $l$ to the entanglement cut and calculate the EE as a function of time for the state's subsequent time evolution. We find that the excess EE compared with the groundstate is a monotonically increasing function which is vanishingly small before the onset at $t \sim l^2$ and eventually saturates to a constant proportional to the scaling dimension of the vertex operator. The quasi-particle picture can interpret the final saturation as the exhaustion of the quasi-particle pairs, while the diffusive nature of the time scale $t \sim l^2$ replaces the common causality constraint in CFT calculation. To further understand this property, we compute the excess EE of a small disk probe far from the excitation point and find chromatography pattern in EE generated by quasi-particles of different propagation speeds. 
\end{abstract}

\maketitle
\section{Introduction}

In the past two decades, much effort was devoted to the study of the entanglement entropy(EE) in the quantum many-body systems. In the condensed matter systems in particular, EE serves as an valuable probe that can extract information such as topological data\cite{levin_detecting_2006,kitaev_topological_2006} and universal quantities\cite{vidal_entanglement_2003, fradkin_entanglement_2006, hsu_universal_2009, stephan_shannon_2009} at critical points. For a review, see a list of literatures\cite{calabrese_entanglement_2004, calabrese_entanglement_2009, calabrese_quantum_2016, fendley_topological_2007, nishioka_holographic_2009} and references therein. 

The dynamical behavior of the EE is one topic in this field. It provides a route to experimental measurement of EE in an extended system (see \cite{cardy_measuring_2011} for a proposal of measuring R\'enyi entropy in an extended condensed matter system), which is still very difficult up to now (see \cite{islam_measuring_2015} for a static measurement of second Re\'nyi entropy in an optical lattice with 4 cold atoms). The quench dynamics is also theoretically interesting on its own right. In a quench process, one prepares an initial state which is usually the groundstate of some Hamiltonian $H$ and then unitarily evolves the system with a \emph{different} Hamiltonian $H'$. The EE is presumed to capture the spreading of excitation on the way to equilibrium. 

There are generally two types of quench protocols:
\begin{enumerate}
\item Global quench. The global quench protocol evolves the ground state of $H$ with a globally different Hamiltonian $H'$. Consequently, the initial state has an extensive amount of extra energy compared to the ground state of $H'$ and EE will grow tremendously afterwards. Computation in translational invariant 1+1d critical systems shows a linear growth of EE immediately after applying the new Hamiltonian. The EE ultimately saturates to a value that is proportional to the size of the subsystem\cite{calabrese_entanglement_2007}.
\item Local quench. In contrast, the local quench protocol only weakly perturbs the initial system. One example of this (usually called ``local quench'' or ``cut and join'' protocol in the literature) is to prepare two identical 1d ground states of $H$ and then join them together and evolve with the same form Hamiltonian of the doubled system. The only difference lies at the connecting points where the dynamics that used to be set by boundary conditions is now determined by a bulk term in the Hamiltonian of the doubled system. Again in the critical system, a CFT calculation reveals a logarithmic growth of EE whose coefficients is the 1/3 of the central charge of the CFT. This result is similar to the EE of a single interval on ground state of 1+1s CFT, in which case the length of the interval is in place of the time difference here. 
\end{enumerate}

What we will study in this paper is a local quench weaker than ``cut and join'', which is called quench of local operator excitations. As the name suggests, we let a local operator act on the initial state and then evolve with the same Hamiltonian. Equivalently, one sets the new Hamiltonian $H'$ as the sum of $H$ and a delta function pulse of local operators at the moment just before the quench. One meaningful measure here is the excess EE compared to that of the ground state, which is expected to reflect the strength and spreading of excitation created by the local operators. In critical systems described by a CFT, local primary field excitations are studied in \cite{he_quantum_2014} and the excess EE increases for a long time to a limiting value equal to the logarithm of the quantum dimension\cite{he_quantum_2014}. The growth of excess EE after the local operator excitation is constrained by causality: the excess EE is zero until the signal traveling at speed of light reaches the entanglement cut. The causality constraint and the saturation behaviors are kept in free boson systems in higher dimensions\cite{nozaki_quantum_2014}; the only change is the way of development of excess EE from zero to its maximal value.

The results of all these three protocols can be quantitatively explained by the physical picture of quasi-particles. The extra energy compared to the ground state of $H'$ is assumed to be carried by coherent quasi-particle pairs. The subsequent time evolution separates the individual quasi-particles and EE is gained when the one of them in the pairs crosses the entanglement cut. In short, the generation of excess EE is ascribed to the proliferation and propagation of those quasi-particle pairs. We give a review about this framework in section \ref{sec:quasi}.

As far as we know, there is yet no analytic result of EE of local operator excitations in a non-relativistic system. Consequently, in this paper, we study the excess EE in such a system. The one we study is the quantum Lifshitz model whose dynamical exponent $z = 2$ (while CFT has $z = 1$) in the presence of the local vertex operator excitation. The model describes a critical line of the quantum eight-vertex model with one special point corresponding to the quantum dimer model on bipartite lattice. The scale invariance of the ground state wavefunction is what makes analytic calculation possible.

We take two subsystems, one the upper half plane, the other a disk and find that the excess EE will grow immediately after the local excitation and reach a limiting value of order the scaling dimension of the vertex operator. The typical time scale when the excess EE is considerable, i.e. of order of the maximal value, is the distance from excitation to the entanglement cut \emph{squared}. This is when the quasi-particles diffuse to the entanglement cut, in consistent with $z = 2$. We also find small plateau structures in the short time dynamics of excess EE and conjecture that it reveals quasi-particle density of states and possible dispersion of different species during the propagation. In summary, the quasi-particle picture can still qualitatively interpret the results with a slight modification that replaces causality constraint with a diffusive behavior. 

The structure of this paper is as follows. In section \ref{sec:quasi}, we review the quasi-particle picture of interpreting the dynamics of EE in the two quench protocols. Then in section \ref{sec:intro_qlifhz}, we introduce the quantum Lifshitz model and the vertex operator excitation we will be focusing on. We define the excess EE in section \ref{sec:excess_EE} and evaluate it for the upper half plane and disk in section \ref{sec:result}. We finally summarize our results in section \ref{sec:summary}.

\section{Review of the quasi-particle picture}
\label{sec:quasi}

Quasi-particle picture is a heuristic way of understanding the phenomenology of EE change in the quench problem. It is exact in the CFT calculation, and is also believed to be valid beyond CFT. Here we review the quasi-particle interpretation of the two quench protocols in order. 

\begin{figure}[h]
\begin{minipage}[t]{0.47\linewidth}
\centering
\includegraphics[width=\textwidth]{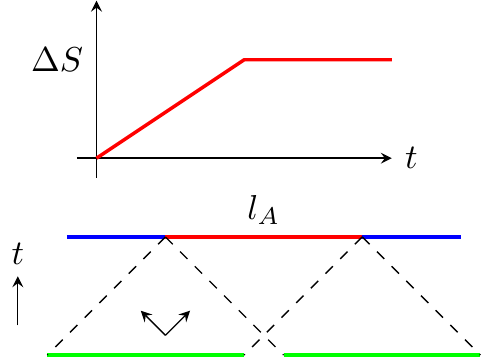}
\caption{Quasi-particle picture for global quench protocol. This is a space time diagram where at time $t$, region A and B are labelled by red and blue lines. The coherent pair of quasi-particles are generated uniformly on each point and radiated in the direction of light cone. The green region encloses sites where part of the quasi-particle pair is in region A at time $t$. The length of green region grows linearly and saturates to value $l_A$ after $t = \frac{l_A}{2}$. }
\label{fig_gq_quasi.pdf}
\end{minipage}
\hfill
\begin{minipage}[t]{0.47\linewidth}
\centering
\includegraphics[width=0.8\textwidth]{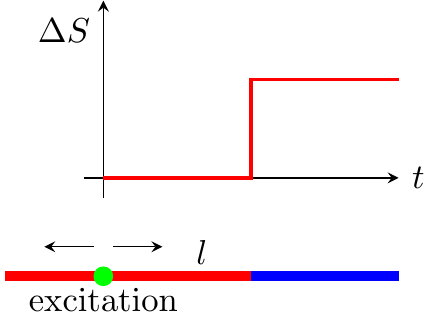}
\caption{Quasi-particle picture for local primary field excitation in 1+1d CFT. Figure below shows A and B to the two semi-infinite systems. The local excitation is located at a distance $l$ to the entanglement cut. The excess EE remains zero before quasi-particle's arrival and burst into log of quantum dimension afterwards. }
\label{fig_lq_quasi.pdf}
\end{minipage}
\end{figure}

In the global quench protocol, the excess energy compared with the true ground state of the Hamiltonian $H'$ is distributed across the system. For a translational invariant Hamiltonian, same types of quasi-particles are radiated on each point of the system. The number of entangled pair between region A and B is proportional to the area of the green region in Fig. \ref{fig_gq_quasi.pdf}, which grows linearly and saturates to the maximal value after $t > \frac{l_A}{2}$ ($l_A$ is the length of subsystem A). This explains the linear growth and extensive saturation value\cite{calabrese_evolution_2005,calabrese_quantum_2016}.

In the ``cut-and-join'' protocol of local quench, the extra energy is only distributed in the vicinity of the joint points. If we choose the region A to be a single interval that has distance $l$ to the joint point, then the EE will keep the ground state value of $ \frac{c}{3} \ln l_A + \text{constant}$ until the time $\frac{l_A}{c}$ when the quasi-particle traveling at the speed of light arrives the entanglement cut\cite{calabrese_entanglement_2007,calabrese_quantum_2016}. It will grow logarithmically afterwards. This picture naturally gives rise to the horizon effect. 

For the quench of local operator excitation, the quasi-particle is created exactly at the point of the operator insertion. Again if the excitation point has a distance to the entanglement cut, causality constraint forces the entanglement to be unchanged until the arrival of quasi-particles, see Fig. \ref{fig_lq_quasi.pdf}. Here the EE will not be extensive since the local excitation only add a very small amount of single particle energy to the ground state. Its strength can be quantified by the quantum dimension, which represents the degrees of freedom of the quasi-particles. We see that the saturated value of excess EE is indeed proportional to this strength\cite{he_quantum_2014}. This is another example of extracting topological data from EE. 

\section{Introduction to Quantum Lifshitz Model and Its Dynamics}
\label{sec:intro_qlifhz}
\subsection{Quantum Hamiltonian}
The Quantum Lifshitz model is a compact boson theory that describes the critical behavior of the quantum eight-vertex model\cite{ardonne_topological_2004,fradkin_field_2013}. The quantum Lifshitz model has Hamiltonian
\begin{equation}
\label{eq:qlif-H}
H = \int d^2x \, \frac{1}{2} \Big\{ \Pi^2+   g^2  \big[\nabla^2\phi\big]^2 \Big\} \, ,
\end{equation}
where $\phi$ is a compact boson field $\phi \sim \phi + 2\pi R_c$ and $\Pi = \dot{\phi}$ is its conjugate momentum. Due to the absence of the regular stiffness term $(\nabla \phi)^2$, this theory does not have Lorentz symmetry. The dynamic exponent $z$ is $2$.

By varying values of $g$, the Hamiltonian in equation \eqref{eq:qlif-H} can in general model a critical line of the quantum eight vertex model\cite{ardonne_topological_2004}. What we have in mind however, is the Rokhsar-Kivelson(RK) critical point of the square lattice quantum dimer model\cite{moessner_quantum_2011,fradkin_entanglement_2006,henley_relaxation_1997,ardonne_topological_2004} which is at $g = \frac{1}{8\pi}$. There, the compact boson field is naturally identified as the coarse grained height field on square lattice\cite{ardonne_topological_2004,moessner_quantum_2011}.

It is generally believed that the Hamiltonian \eqref{eq:qlif-H} gives the correct time evolution of the quantum dimer model. We here present two heuristic ways to justify this point.

One of the them is a Ginzburg-Landau type argument that keeps lowest order possible terms that consistent with the required symmetry\cite{fradkin_field_2013}. In the dimer problem, translational and rotational symmetries enforce the Hamiltonian to have the following form
\begin{equation}
H_0 = \int d^2x \, \frac{1}{2} \Big\{ \Pi^2+  A (\nabla \phi)^2 +  g^2  \big[\nabla^2\phi\big]^2 \Big\} .
\end{equation}
When $A > 0 $, the system will flow to a phase that pin the $\phi$ field to fixed value, which is identified to be the columnar phase away from the RK point. On the other hand, $A < 0$ corresponds to an unstable Hamiltonian that is not semi-positive definite. At $A = 0$, $H_0$ reduces to the Hamiltonian \eqref{eq:qlif-H} and can be diagonalized using
\begin{equation}
Q(\vec{x}) = \frac{\delta}{\delta \phi( \vec{x} )} - g \nabla^2 \phi(\vec{x} )  \qquad Q^{\dagger}(\vec{x}) = - \frac{\delta}{\delta \phi( \vec{x} )} - g \nabla^2 \phi(\vec{x} )
\end{equation}
as a series of harmonic oscillators (normal ordered), such that
\begin{equation}
\label{eq:H-normal-order}
H = \int d^2 x \, Q^{\dagger} (\vec{x}) Q( \vec{x} )
\end{equation}
The ground state wavefunction is thus annihilated by $Q(\vec{x})$ and has a Gaussian form
\begin{equation}
|\text{gnd} \rangle = \frac{1}{\sqrt{\mathcal{Z}}} \int [d\phi] \exp\big[ - \frac{g}{2}  \int d^2x  \nabla \phi \cdot \nabla \phi \big] | \phi \rangle 
\end{equation}
where $\mathcal{Z}$ is partition function of the free compact Boson
\begin{equation}
\mathcal{Z} = \int [d\phi] \exp\big[ - g \int d^2x  \nabla \phi \cdot \nabla \phi \big]
\end{equation}
This reproduces the fact that at Rokhsar-Kivelson critical point, the dimer density operator (derivative of the boson)\cite{kasteleyn_statistics_1961,fisher_statistical_1961} has a power law correlation function.

Another independent derivation is proposed by Nienhuis\cite{nienhuis_two_1987} and Henley\cite{henley_relaxation_1997} who map the quantum Hamiltonian of ``flipping'' dynamics to a Monte Carlo process. The classical Monte Carlo process gives exactly the same probability distribution as the ground state wavefunction on the dimer basis. And the relaxation to the equilibrium is the imaginary time quantum evolution. To obtain a continuous description, the stochastic process is heuristically written as a Langevin equation with a Gaussian noise. The Hamiltonian \eqref{eq:H-normal-order} which governs the corresponding master equation is then identified as the effective Hamiltonian. 

The vertex operator $e^{i \alpha \phi}$ ($\alpha 2\pi R_c \in \mathbb{Z}$) is a sensible set local operators in the compact boson theory which create a Boson coherent state. It is also the electric operator in the quantum dimer model context\cite{fradkin_field_2013}. We consider the excess EE generated by this local excitation. Specifically, we act the vertex operator $e^{i \alpha \phi}$ on the ground state and evolve for time $t$ to reach a state
\begin{equation}
|\vec{x}, t \rangle = e^{- i H t} e^{i \alpha  \phi( \vec{x})  } | \text{gnd}\rangle.
\end{equation}
The excess EE is defined to be the difference of EE between $| \x, t \rangle$ and $| \x , 0 \rangle$. As a function of time, it should reflect the spreading of the local excitation.

\subsection{Time Evolution of the Vertex Operator}

In this subsection, we solve the time evolution equation and express the $|\x,t\rangle$ in terms of the ground state boson operators. For clarity, in this subsection we temporarily turn to the hatted notation $\hat{\phi}$ to denote the boson operator and reserve the unhatted $\phi$ for the eigenvalue of the $|\phi \rangle$ basis. 

Since the ground state is annihilated by $H$, we rewrite the state $|\x,t\rangle$ as
\begin{equation}
| \vec{x},t \rangle = e^{- i \hat{H} t } e^{i \alpha \hat{ \phi} ( \vec{x} ) } e^{i \hat{H} t } | \text{gnd} \rangle  = e^{ i \alpha \hat{\phi}( \vec{x}, -t )} | \text{ gnd} \rangle 
\end{equation}
where the time dependent boson operator $\hat{ \phi} ( \vec{x}, -t ) $ is the solution of Heisenberg equation
\begin{equation}
i \frac{\partial \hat{\phi}( \vec{x},-t ) }{\partial t } =  [ \hat{H} , \hat{\phi} ( \vec{x}, -t ) ] = e^{- i \hat{H} t}[\hat{H}, \hat{\phi}(\vec{x}) ] e^{i \hat{H} t} = [ Q^{\dagger}( \vec{x}, -t ) - Q( \vec{x} , -t ) ]
\end{equation}
The $-$ sign in $t$ indicates another equivalent convention used in \cite{he_quantum_2014} that interprets the $-t$ as the time of local excitation and $0$ as the time of measurement. 

The $Q^{\dagger}$ and $Q$ are the non-standard creation/annihilation operators; they have the commutation relation
\begin{equation}
  [Q(\vec{x}), Q^{\dagger}(\vec{y})] =  -2g \nabla^2_x \delta ( \vec{x} - \vec{y}).
\end{equation}
We can then solve their Heisenberg equations
\begin{equation}
\begin{aligned}
  \partial_t Q^{\dagger} &= i [ H, Q^{\dagger}] = -i 2g \nabla^2 Q^{\dagger} \\
  \implies \quad  & Q^{\dagger}(\vec{x}, t) = e^{-i 2g t\nabla^2 }Q^{\dagger}(\vec{x})
\end{aligned}
\end{equation}
which gives the time evolution of the boson operator
\begin{equation}
\begin{aligned}
\hat{\phi}(\vec{x}, -t ) &= -i \int^t \big[e^{2igs \nabla^2} Q^{\dagger}(\vec{x}) -e^{-2igs \nabla^2} Q(\vec{x}) \big]ds 
\end{aligned}
\end{equation}

Now consider acting the operator $Q$ on the ground state. We have
\begin{equation}
Q( \vec{x} ) |\text{gnd} \rangle  = 0 \qquad Q^{\dagger}( \vec{x} ) |\text{gnd} \rangle  = -2g \nabla^2 \hat{\phi}( \vec{x} ) |\text{gnd} \rangle
\end{equation}
and hence
\begin{equation}
\hat{\phi}( \vec{x}, -t ) |\text{gnd} \rangle = \int^t ds\, e^{2igs \nabla^2} (2ig \nabla^2 )\hat{\phi}(\vec{x}) |\text{gnd}\rangle  = e^{2igt \nabla^2}\hat{\phi}(\vec{x}) | \text{gnd}\rangle
\end{equation}

One should interpret $e^{2i gt \nabla^2}$ as a Sch\"odinger time evolution. We add a small real positive constant $\epsilon$ to the imaginary time
\begin{equation}
\tau = \epsilon + 2igt 
\end{equation}
to control the ultraviolet divergence; this is the damping term that screens out the high energy mode\cite{calabrese_evolution_2005}. The operator $e^{ \tau \nabla^2} \hat{\phi}( \vec{x} )$ obeys
\begin{equation}
\partial_{\tau} \big[ e^{ \tau \nabla^2} \hat{\phi} ( \vec{x} )\big] = \nabla^2 \big[e^{ \tau \nabla^2} \hat{\phi} ( \vec{x} )\big]
\end{equation}
whose solution in free space is given by
\begin{equation}
e^{ \tau \nabla^2} \hat{\phi} ( \vec{x} ) = \hat{\phi}( \vec{x}, \tau) = \int d^2x \, H( \vec{x}, \tau; \vec{x}' ) \hat{\phi}(\vec{x}')
\end{equation}
where $H$ is the standard heat kernel in 2d
\begin{equation}
  H(\vec{x}, \tau; \vec{x}' ) = \frac{1}{4\pi\tau} \exp\left\{ - \frac{(\vec{x}-\vec{x}')^2}{4 \tau} \right\}
\end{equation}

The time evolved vertex operator is thus
\begin{equation}
e^{i \alpha \hat{\phi}( \vec{x}, -t ) } = \exp\big\{ i \alpha e^{ \tau \nabla^2} \hat{\phi} ( \vec{x} )  \big\}
\end{equation}
on the ground state.

\section{Excess Entanglement Entropy}
\label{sec:excess_EE}
In this section, we define and derive the replica formula for the excess EE. 

The time dependent EE after the local excitation at $t = 0 $ is the Von Neumann entropy
\begin{equation}
S( t) = -\text{tr}[ \rho_A(t) \ln \rho_A( t)  ]
\end{equation}
w.r.t the time dependent density matrix 
\begin{equation}
  \rho_A (t) = \text{tr}_B[ \rho ] = \text{tr}_B \big[ | \vec{x}, t \rangle \langle \vec{x}, t |  \big]
\end{equation}
associated with the state $|\x,t\rangle$.

The excess EE is defined to be
\begin{equation}
\Delta S( t ) = S(t ) - S( 0 ) 
\end{equation}
where the $\Delta$ symbol in this paper always denote the difference between time $t$ and time 0. 

The way to compute EE is to take the analytic continuation of the R\'enyi EE
\begin{equation}
S_n = -\frac{1}{n-1}\ln \text{tr}\rho_A^n 
\end{equation}
at $n = 1$. In field theory setting, the quantity $\text{tr}\rho_A^n$ has $n$ replicated fields properly glued together\cite{calabrese_entanglement_2009}, thus the name ``replica trick''. 

We evaluate this time dependent EE by using replica trick at each time slice. In fact, the EE of static ground state EE has been evaluated and refined by many groups \cite{fradkin_entanglement_2006,oshikawa_boundary_2010,zaletel_logarithmic_2011,stephan_shannon_2009,zhou_entanglement_2016}. Here we extend the trick used in \cite{fradkin_entanglement_2006,zhou_entanglement_2016} to an excited state. 

\begin{figure}[h]
\centering
\includegraphics[width=1\textwidth]{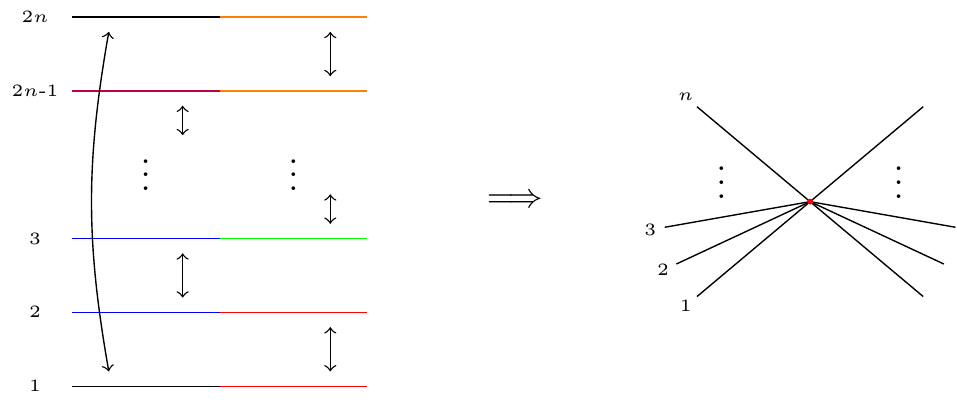}
\caption{The gluing conditions for the fields of even indices. The left figure shows a cyclic relation that neighboring copies have the same value(same color in the figure) in one of the subsystems. The right figure shows the collapsing of odd fields to form $n$ independent copies that share the same value \emph{only} on the entanglement cut.}
\label{fig:gluing}
\end{figure}

The extension is different from the replica trick commonly used in CFT calculation\cite{calabrese_entanglement_2009}, so we give a brief review of its derivation. We use discrete notation to derive $\text{tr}(\rho_A^n)$. Denoting $|a\rangle$ and $|b \rangle$ as sets of complete orthonormal basis on subsystem A and B, for the ground state density matrix $\rho = |\text{gnd}\rangle\langle\text{gnd}|$, we have
\begin{equation}
\begin{aligned}
\text{tr}(\rho_A^n) & = \sum_{b_i}\text{tr}\big[ \langle b_1 |\rho | b_2 \rangle \delta_{b_1 b_2} \cdots \langle b_{2n-1} |\rho | b_{2n} \rangle \delta_{b_{2n-1} b_{2n}} \big]\\
&= \sum_{a_i, b_i}\langle a_1 | \langle b_1 |\rho | b_2 \rangle |a_2\rangle  \cdots \langle a_{2n-1} | \langle b_{2n-1} |\rho | b_{2n} \rangle |a_{2n}\rangle \\
&\quad\quad   \delta_{a_1 a_{2n}} \prod_{i=1}^{n-1}\delta_{a_{2i} a_{2i+1}} \prod_{i=1}^{n}\delta_{b_{2i-1} b_{2i} } \\
&\propto \sum_{a_i,b_i} \exp\Big\{- \sum_{i=1}^{2n} \frac{1}{2}S[\phi_i] \Big\} \delta_{a_1 a_{2n}} \prod_{i=1}^{n-1}\delta_{a_{2i} a_{2i+1}} \prod_{i=1}^{n}\delta_{b_{2i-1} b_{2i} }.
\end{aligned}
\end{equation}
This expression has $2n$ copies of fields, while the delta functions enforce the constraint that the fields of odd indices are created by concatenating parts from adjacent fields of even indices. It is then equivalent to remove those odd fields; meanwhile duplicate the even fields and require them to have same value on the cut. The later condition on cut ensures the possibility of stitching two parts from two even fields to create the odd field between them. This procedure is depicted in \ref{fig:gluing}

The derivation for the excited state is almost the same except for the insertion of operators $O(\phi,t) = \exp( i \alpha e^{\tau \nabla^2 } \phi ( \vec{x} ) )$ and $O^{\dagger}(\phi,t) = \exp( - i \alpha e^{-\tau \nabla^2 } \phi^{\dagger} ( \vec{x} ) )$ in front of the partition function. 
\begin{equation}
\text{tr}(\rho_A^n) \propto \sum_{a_i,b_i} \prod_{i=1}^{n} O( \phi_{2i-1}, t ) O^{\dagger}( \phi_{2i}, t ) \exp\Big\{- \sum_{i=1}^{2n} \frac{1}{2}S[\phi_i] \Big\} \delta_{a_1 a_{2n}} \prod_{i=1}^{n-1}\delta_{a_{2i} a_{2i+1}} \prod_{i=1}^{n}\delta_{b_{2i-1} b_{2i} }.
\end{equation}
So relabeling the surviving even field from $1$ to $n$, we have
\begin{equation}
\text{tr}(\rho_A^n) = \frac{\langle \prod_{i=1}^n O( \phi_i, t ) O^{\dagger}( \phi_i, t )  \rangle_{\text{glue}} }{\langle  O( \phi, t ) O^{\dagger}( \phi, t )  \rangle^n_{\rm Free} }
\end{equation}
where the $2n$ point function in the numerator is evaluated on the manifold in Figure \ref{fig:gluing}. This formula has similar structure as the CFT calculation in \cite{he_quantum_2014}, where the 2$n$ point function in the numerator is evaluated on the $n$-sheeted Riemann surface. 

We use target space rotation to deal with the gluing condition. First we separate the field into classical and quantum part
\begin{equation}
\phi = \varphi + \phi_{\text{cl}} 
\end{equation}
such that $\phi_{\rm cl}( \x )\big|_{\rm cut} = \phi( \x ) \big|_{\rm cut}$ and $\nabla^2 \phi_{\text{cl}} = 0$, then the quantum fluctuation has Dirichlet boundary condition $\varphi( \x )\big|_{\text{cut}} = 0$ and the action separates 
\begin{equation}
S[\phi] = S[\varphi] + S[\phi_{\text{cl}}] 
\end{equation}
Notice that the classical part does not evolve with time
\begin{equation}
O( \phi, t ) = \exp \big(i \alpha e^{ \tau \nabla^2 \phi( \vec{x} ) } \big)  = \exp \big(i \alpha e^{ \tau \nabla^2 \varphi( \vec{x} ) } \big) \exp( i \alpha \phi_{\text{cl}} ) = O( \varphi, t ) \exp( i \alpha \phi_{\text{cl}} ), 
\end{equation}
so in the average on the glued manifold $\langle \cdots \rangle_{\text{glue}} $, the real valued classical field on the vertex operator cancels, 
\begin{equation}
\label{eq:2pt-separate}
\begin{aligned}
\langle \cdots \rangle_{\text{glue}} &= \langle \cdots \rangle_{\varphi} \sum_{\phi_{\rm cl}^i} \exp\Big[ i \alpha \sum_{j=1}^n (\phi^j_{\text{cl}} - \phi^{j \dagger}_{\text{cl}} ) - \sum_{i=1}^n S[\phi^i_{\text{cl}} ]\Big] \\
&=\langle \cdots \rangle_{\varphi} \sum_{ \phi_{\rm cl}^i } \exp\Big[ - \sum_{i=1}^n S[\phi^i_{\text{cl}} ]\Big]
\end{aligned}
\end{equation}
where the summation is subject to the boundary condition
\begin{equation}
\phi_{\text{cl}}^1(\x) = \phi_{\text{cl}}^2(\x) =\cdots =   \phi_{\text{cl}}^n(\x) = \phi_{\text{cut} }(\x) \qquad \mod \quad  2\pi R_c  \,\, \text{on cut} 
\end{equation}
Since the $\varphi$ field satisfies Dirichlet boundary condition on the entanglement cut, we have 
\begin{equation}
\langle \cdots  \rangle_{\varphi} = \langle O( \varphi, t ) O^{\dagger}( \varphi, t )  \rangle^{n}_{\text{Dirichlet}}
\end{equation}
as a result 
\begin{equation}
\langle \cdots \rangle_{\text{glue}} 
=\langle O( \varphi, t ) O^{\dagger}( \varphi, t )  \rangle^{n}_{\text{Dirichlet}} \sum_{ \phi_{\rm cl}^i } \exp\Big[ - \sum_{i=1}^n S[\phi^i_{\text{cl}} ]\Big]
\end{equation}
The summation of the classical modes is the same as the ground state, so that rotational trick used there works in the same way. We rotate these classical fields by the following unitary matrix
\begin{equation}
\begin{bmatrix}
\bar{\phi}_{\text{cl}}^1 \\
\bar{\phi}_{\text{cl}}^2 \\
\bar{\phi}_{\text{cl}}^3 \\
\cdots\\
\bar{\phi}_{\text{cl}}^{n-1} \\
\bar{\phi}_{\text{cl}}^{n}\\
\end{bmatrix}
=
\begin{bmatrix}
\frac{1}{\sqrt{2}} & \frac{-1}{\sqrt{2}} & & & &\\
\frac{1}{\sqrt{6}} & \frac{1}{\sqrt{6}} & \frac{-2}{\sqrt{6}} & & &\\
\frac{1}{\sqrt{12}} & \frac{1}{\sqrt{12}} & \frac{1}{\sqrt{12}} &  \frac{-3}{\sqrt{12}}& &\\
\cdots & \cdots & \cdots & \cdots & \cdots & \\
\frac{1}{\sqrt{n^2 -n}} & \frac{1}{\sqrt{n^2 -n}} & \cdots & \cdots  & \frac{-(n-1)}{\sqrt{n^2 -n}}&\\
\frac{1}{\sqrt{n}}& \frac{1}{\sqrt{n}}& \frac{1}{\sqrt{n}}& \cdots& \frac{1}{\sqrt{n}}& \\
\end{bmatrix}
\begin{bmatrix}
\phi_{\text{cl}}^1 \\
\phi_{\text{cl}}^2 \\
\phi_{\text{cl}}^3 \\
\cdots\\
\phi_{\text{cl}}^{n-1} \\
\phi_{\text{cl}}^{n}\\
\end{bmatrix},
\end{equation}
such that after the rotation
\begin{equation}
\bar{\phi}_{\text{cl}}^j( \x)  = 2\pi w_j R_c   \quad \text{ on cut for }  j < n, \quad w_j \in \mathbb{Z}
\end{equation}
It is noted that the $n$-th field is the center of mass mode that is the only one dependent on the value on the cut
\begin{equation}
\bar{\phi}_{\text{cl}}^{n}(\x)\big|_{\text{cut}} = \sqrt{n} \phi_{\text{cut}}(\x)  
\end{equation}
while the rest of the fields decouple
\begin{equation}
\sum_{\phi_{\rm cl}^i} \exp\Big[ - \sum_{i=1}^n S[\phi^i_{\text{cl}} ]\Big] = \bigg\{ \sum_{\phi_{\text{cut}}} \exp\Big[ - S [\bar{\phi}^n_{\text{cl}} ] \Big]\bigg\}  \bigg\{\sum_{\vec{w}\in \mathbb{Z}^{n-1} }  \exp\Big[ - \sum_{i=1}^{n-1} S[\bar{\phi}^i_{\text{cl}} ]\Big] \bigg\}
\end{equation}

The degrees of freedom on the entanglement cut determine $\bar{\phi}^n_{\rm cl}$, while the rest determines the Dirichlet two point function, hence if we combine the two, we should collect all the degrees of freedom in this region and end up with a free two point function 
\begin{equation}
\begin{aligned}
\langle O( \varphi_n, t ) O^{\dagger}( \varphi_n, t )  \rangle_{\text{Dirichlet}} \sum_{\phi_{\rm cut}}  \exp[ - S[\phi^n_{\text{cl}} ]]
& = \langle O( \varphi_n, t ) O^{\dagger}( \varphi_n, t )  \rangle_{\text{Free}} \\
& = \langle O( \phi_n, t ) O^{\dagger}( \phi_n, t )  \rangle_{\text{Free}} 
\end{aligned}
\end{equation}
In fact, we can do this with the identification of the field $\phi_n$
\begin{equation}
\phi_n= \varphi_n + \bar{\phi}_{\text{cl}}^n
\end{equation}

This free two point function will cancel one of the two point functions in the denominator. We therefore have
\begin{equation}
\text{tr}(\rho_A^n) = \bigg\{\frac{\langle  O( \varphi, t ) O^{\dagger}( \varphi, t )  \rangle _{\text{Dirichlet}} }{\langle  O( \varphi, t ) O^{\dagger}( \varphi, t )  \rangle_{\rm Free} } \bigg\}^{n-1} \bigg\{\sum_{\vec{w}\in \mathbb{Z}^{n-1} }  \exp\Big[ - \sum_{i=1}^{n-1} S[\phi^i_{\text{cl}} ]\Big] \bigg\}
\end{equation}
Note that the sum over $\phi_{\text{cl}} $ is time independent and as a result will be canceled in the excess EE. The excess R\'enyi entropy is therefore
\begin{equation}
\label{eq:excess-renyi}
\Delta S_n = \Delta \Big\{ \ln \langle  O( \phi, t ) O^{\dagger}( \phi, t )  \rangle_{\rm Free}  - \ln \langle  O( \phi, t ) O^{\dagger}( \phi, t )  \rangle _{\text{Dirichlet}}  \Big\}
\end{equation}
where $\phi$ now denotes the \emph{non-compact} free boson and $\Delta$ denotes the difference between time $t$ and $0$.

\subsection{Green Function}
\label{subsec:green-fun}
The two point function of the vertex operators ultimately will be reduced to the Green function of the boson field. In this subsection, we define and calculate the free space Green function in space and time directions.

The equal time Green function on the ground state
\begin{equation}
G( \vec{x}_1 , \vec{x}_2 ) = \langle \phi(\vec{x}_1) \phi( \vec{x}_2 )  \rangle
\end{equation}
satisfies Laplace equation
\begin{equation}
- 2g \nabla_1^2 G( \vec{x}_1 , \vec{x}_2 ) = \delta( \vec{x}_1 - \vec{x}_2 ) 
\end{equation}

In free space($\mathbb{R}^2$ plane), the solution is well known
\begin{equation}
  G( \vec{x}_1, \vec{x}_2 ) = -\frac{1}{4\pi g} \ln |\vec{x}_1 - \vec{x}_2 |
\end{equation}

We define the equal space Green function, which is useful later, to be the Green function evaluated at the same position but different imaginary time
\begin{equation}
\begin{aligned}
G( \tau_1, \tau_2 ) &=   \langle e^{ \tau_1 \nabla^2 }\phi( \vec{x} ) e^{ \tau_2 \nabla^2 } \phi^{\dagger} ( \vec{x} ) \rangle  \\ 
&=\int d^2x_1 d^2x_2 \, H(\vec{x}, \tau_1; \vec{x}_1)  G( \vec{x}_1 , \vec{x}_2 )  H( \vec{x}, \tau_2; \vec{x}_2 )  \\
\end{aligned}
\end{equation}
By taking the $\tau_1$ derivative and use the property of the heat kernel,
\begin{equation}
\begin{aligned}
-2g \partial_{\tau_1} G( \tau_1, \tau_2 ) &= \int d^2x_1d^2x_2 \, H(\vec{x}, \tau_1; \vec{x}_1)(-2g \nabla_1^2 )  G( \vec{x}_1 , \vec{x}_2 )  H( \vec{x}, \tau_2; \vec{x}_2 )  \\
&= \int d^2x_1d^2x_2 \, H(\vec{x}, \tau_1; \vec{x}_1)\delta ( \vec{x}_1 , \vec{x}_2 )  H( \vec{x}, \tau_2; \vec{x}_2 )  \\
&= \frac{1}{4\pi (\tau_1 + \tau_2 )}
\end{aligned}
\end{equation}
Convergence of the integral requires $\text{Re}(\tau_1 + \tau_2 ) > 0 $, which is satisfied by adding a damping parameter $\epsilon$ to both imaginary time $\tau_1 $ and $\tau_2$. Up to a constant
\begin{equation}
G( \tau_1, \tau_2 )  = - \frac{1}{8 \pi g} \ln | \tau_1 + \tau_2 | 
\end{equation}
The $\frac{1}{8\pi g}$ rather than $\frac{1}{4\pi g}$ factor is a manifestation of $z = 2$. These two limits of the Green function agree with the general Green function expression in \cite{fradkin_field_2013}. 

The Green function in this case is associated with the operator $-2g \nabla^2$. In the following we will instead calculate  in terms of the Green function $G_{\Delta}( \x_1, \x_2 ) $ of the standard Laplacian operator $-\nabla^2$, and relate it to the two point function via
\begin{equation}
  G_{\Delta} ( \vec{x}_1, \vec{x}_2 )  = 2g G( \vec{x}_1, \vec{x}_2 ) \qquad G_{\Delta} ( \tau_1, \tau_2 )  = 2g G( \tau_1, \tau_2 ) 
\end{equation}

\subsection{Excess Entanglement Entropy in terms of Green Function}

In equation \eqref{eq:excess-renyi}, we need the two point function of the diffused vertex operator
\begin{equation}
  \left\langle \exp\Big\{ i \alpha \phi( \vec{x}, \tau_1 )  \Big\} \exp\Big\{ i \alpha \phi^{\dagger} ( \vec{x}, \tau_2 ) \Big\}  \right\rangle = \left\langle \exp\Big\{ i \alpha [ \phi( \vec{x}, \tau_1 ) - \phi^{\dagger} ( \vec{x}, \tau_2 )] \Big\} \right\rangle
\end{equation}
where
\begin{equation}
\tau_1 = \epsilon + i 2gt \qquad \tau_2 = \epsilon - i 2g t 
\end{equation}
The exponent is a source term in the Gaussian path integral 
\begin{equation}
\begin{aligned}
i \alpha [ \phi( \vec{x}, \tau_1 ) - \phi^{\dagger} ( \vec{x}, \tau_2 )]&= i \alpha \big( e^{i 2g t\nabla^2 } - e^{-i 2g t \nabla^2 } \big)\phi(\x ) \\
&= i \alpha \int d^2x' \big[H( \vec{x}, \tau_1; \vec{x}' ) - H( \vec{x}, \tau_2; \vec{x}' )\big] \phi( \vec{x}' ) \\
&= \int d^2 x' J( \vec{x}') \phi( \vec{x}' ) 
\end{aligned}
\end{equation}
Here $J(\vec{x}')$ is a real operator because
\begin{equation}
J \sim i \alpha \big( e^{i 2g t\nabla^2 } - e^{-i 2g t \nabla^2 } \big)  = 2 \alpha\sin( -2gt \nabla^2 ) 
\end{equation}
we thus have the standard results for Gaussian integral
\begin{equation}
\left\langle \exp\left[ \int d^2x' J( \x' ) \phi( \x' ) \right] \right\rangle  = \exp\Big\{  \frac{1}{2} \int d^2x_1 d^2x_2 J( \vec{x}_1) G( \vec{x}_1 - \vec{x}_2 ) J( \vec{x}_ 2 ) \Big\}
\end{equation}
and therefore
\begin{equation}
\Delta S_n = \Delta \frac{1}{2}  \int d^2x_1 d^2x_2 J( \vec{x}_1) G( \vec{x}_1 - \vec{x}_2 ) J( \vec{x}_ 2 )\Big|^{A\cup B}_{\text{Dirichlet}} 
\end{equation}
The $JGJ$ integral consists of four Green functions in time direction
\begin{equation}
\int J G J = - \alpha^2 \int ( H - H ) G (H - H) = - \alpha^2 \big[ G(\tau_1, \tau_1 ) - G( \tau_1, \tau_2 ) -  G( \tau_2, \tau_1 )+ G( \tau_2, \tau_2 )\big]
\end{equation}
Define the cross Green function 
\begin{equation}
G_{\Delta}( \tau_1, \tau_2, \times ) = \big[ G_{\Delta} (\tau_1, \tau_1 ) - G_{\Delta}( \tau_1, \tau_2 ) -  G_{\Delta}( \tau_2, \tau_1 )+ G_{\Delta}( \tau_2, \tau_2 )\big]
\end{equation}
the excess R\'enyi entropy is
\begin{equation}
\Delta S_n = - \Delta \Big[\frac{\alpha^2}{4g} G_{\Delta}( \tau_1, \tau_2, \times ) \Big|^{A\cup B}_{\text{Dirichlet}} \Big]
\end{equation}

\section{Results and Discussion}
\label{sec:result}
We have obtained the free space Green function in subsection \ref{subsec:green-fun}, where the equal space Green function is evolved from the equal time Green function. When imposing Dirichlet boundary condition on the cut, we can solve Dirichlet problems in A (and similarly in B)
\begin{equation}
- \nabla_1^2 \GDA = \delta( \x_1 - \x_2 ) \quad \GDA\Big|_{\partial A} = 0
\end{equation}
and then construct the Green function on the whole plane as
\begin{equation}
\label{eq:GD-plane}
\begin{aligned}
G_{\Delta}^{\text{Dirichlet}}( \x_1, \x_2 )  =&  \GDA [\theta(\x_1 \in A ) \theta( \x_2 \in A ) ]  \\ 
&+ G_{\Delta}^B( \x_1, \x_2 ) [\theta(\x_1 \in B ) \theta( \x_2 \in B ) ] 
\end{aligned}
\end{equation}
The step function $\theta$ in Equation \eqref{eq:GD-plane} implement the fact that the Dirichlet boundary condition destroys the correlation \emph{between} two regions while modifying it \emph{within} each region through ``boundary charge''. 

Then the equal space Green function is constructed from the equal time Green function through
\begin{equation}
\label{eq:eq_space_green_AB}
\begin{aligned}
G^{\text{Dirichlet}}_{\Delta}( \tau_1, \tau_2 ) &= \int_{A \cup B \times A \cup B } d^2 x_1 d^2 x_2 \, H(\vec{x}, \tau_1; \vec{x}_1)  G_\Delta^{\text{Dirichlet}}( \vec{x}_1 , \vec{x}_2 )  H( \vec{x}, \tau_2; \vec{x}_2 )  \\ 
&= \int_{A \times A}d^2 x_1 d^2 x_2 \, H(\vec{x}, \tau_1; \vec{x}_1)  G_\Delta^{A}( \vec{x}_1 , \vec{x}_2 )  H( \vec{x}, \tau_2; \vec{x}_2 ) \\
& + \int_{B \times B}d^2 x_1 d^2 x_2 \, H(\vec{x}, \tau_1; \vec{x}_1)  G_\Delta^{B}( \vec{x}_1 , \vec{x}_2 )  H( \vec{x}, \tau_2; \vec{x}_2 ) \\
\end{aligned}
\end{equation}

In electrostatic language, the free space $G_{\Delta}( \tau_1, \tau_2)$ is the potential energy between two Gaussian charge distributions (albeit being imaginary), while $G^{\text{Dirichlet}}_{\Delta}( \tau_1, \tau_2 ) $ is the same thing in the presence of induced boundary charge on the entanglement cut. Hence the difference, about which the EE is concerned, only depends on the boundary charge.

In appendix \ref{app:eq_space_green}, we showed that (the double derivative of) the Dirichlet Green function is solely determined by a boundary integral,
\begin{equation}
\partial_{\tau_2 } \partial_{\tau_1 } G^{\text{A}}_{\Delta}( \tau_1, \tau_2 )\Big|^{\text{no cut}}_{\text{Dirichlet}}  = -\int_{\partial A \times \partial A } \H 1  \partial_{n_1} \partial_{n_2} \GDA  \H 2  \, dl_1 \, dl_2 
\end{equation}

We can integrate the heat kernel once to define
\begin{equation}
f( \x ) = \int^{\tau_1}_0 H( 0, \tau; \x_1 ) \, d \tau 
\end{equation}
and recognize that
\begin{equation}
\bar{f}( \x ) = \int_0^{\tau_2} H( 0, \tau; \x_1 ) d \tau 
\end{equation}
Using this, the cross Green function becomes
\begin{equation}
G_{\Delta}( \tau_1, \tau_2, \times ) \Big|^{A\cup B}_{\text{Dirichlet}} = 4 \int_{\partial A \times \partial A } \text{Im}f(\x_1)  K( \x_1, \x_2 )  \text{Im} f(\x_2 )  \, dl_1 \, dl_2 
\end{equation}
where $K(\x_1, \x_2)$ is the kernel on the boundary 
\begin{equation}
K( \x_1, \x_2 ) = \partial_{n_1} \partial_{n_2} \GDA  
\end{equation}
which is singular when $\x_1$ is approaching $\x_2$. In general we can regularize it as
\begin{equation}
\label{eq:cross-Green}
G_{\Delta}( \tau_1, \tau_2, \times ) \Big|^{A\cup B}_{\text{Dirichlet}} = -2 \int_{\partial A \times \partial A } \Big[\text{Im}f(\x_1)- \text{Im} f(\x_2 )   \Big]^2  K( \x_1, \x_2 )  \, dl_1 \, dl_2 
\end{equation}

The excess Re\'nyi EE is therefore
\begin{equation}
\label{eq:EEE}
\Delta S_n = \frac{\alpha^2}{2g} \int_{\partial A \times \partial A } \Big[\text{Im}f(\x_1)- \text{Im} f(\x_2 )   \Big]^2  K( \x_1, \x_2 )  \, dl_1 \, dl_2 
\end{equation}
In the following, we analyze two cases where the Green function can be easily figured out by methods of images.

\subsection{Infinite Plane}
\subsubsection{Excess EE}

\begin{figure}[h]
\centering
\includegraphics[width=0.2\textwidth]{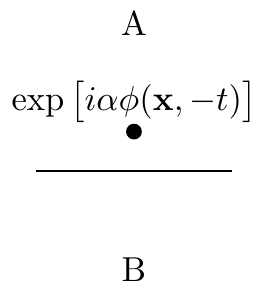}
\caption{The system is an infinite plane. Local operator is placed at $(0, y)$.}
\label{fig:rect}
\end{figure}
We consider the geometry of infinite plane with entanglement cut on the x-axis. In complex coordinate, the equal time Green function in region A can be easily written down
\begin{equation}
\GDA = - \frac{1}{2\pi} \ln |z_1 - z_2 | + \frac{1}{2\pi} \ln | z_1 - \bar{z}_2 | \quad \x_1, \x_2 \in A 
\end{equation}
the kernel
\begin{equation}
\begin{aligned}
K ( \x_1, \x_2 )\Big|_{\partial A} &= \lim_{y_2\rightarrow 0} \partial_{y_2} \partial_{y_1} \GD \Big|_{y_1 = 0 }  = \frac{1}{\pi}\lim_{y_2\rightarrow 0}\partial_{y_2}  \frac{y_2 }{(x_1- x_2 )^2 + y_2^2} 
\end{aligned}
\end{equation}
is indeed singular at $x_1 = x_2$. In appendix \ref{app:distri-bd-int} we provide a way to interpret the distributional integral. The resulting recipe the same as the general formula \eqref{eq:cross-Green}. 

The integral of the heat kernel
\begin{equation}
\text{Im}f(\x_1) = -\frac{1}{4\pi}\text{Si}[\infty] + \frac{1}{4\pi}\text{Si}\Big[\frac{(\x - \x_1)^2}{4t}\Big]
\end{equation}
is given by sin integral function
\begin{equation}
\text{Si}( x) = \int_0^x \sin t^2 \, dt 
\end{equation}

By equation \eqref{eq:EEE}, the excess EE is
\begin{equation}
\begin{aligned}
\Delta S_n =  - \frac{\alpha^2 }{4g}G_{\Delta}^{\text{A}}( \tau_1, \tau_2, \times ) &= \frac{\alpha^2}{32\pi^2g} \int_{\partial A \times \partial A } \frac{\Big\{\text{Si}\big[\frac{x_1^2+y^2}{4t}\big] - \text{Si}\big[\frac{x_2^2+y^2}{4t}\big] \Big\}^2 }{\pi (x_1 - x_2)^2}  \, dx_1 \, dx_2 \\
&= \frac{\alpha^2}{32\pi^2g} \int_{\partial A \times \partial A } \frac{\Big\{\text{Si}\big[x_1^2+\frac{y^2}{4t}\big] - \text{Si}\big[x_2^2+\frac{y^2}{4t}\big] \Big\}^2 }{\pi (x_1 - x_2)^2}  \, dx_1 \, dx_2 \\ 
\end{aligned}
\end{equation}
We can also write the integral in Fourier space through the standard Hilbert transform,
\begin{equation}
\begin{aligned}
\Delta S_n = &= \frac{\alpha^2}{16\pi^2 g} \int_{-\infty}^{\infty} \text{Si}\big[x^2+\frac{y^2}{4t}\big] \partial_x (\text{Si}\big[x^2+\frac{y^2}{4t}\big])_{\mathcal{H}} dx \\
&=\frac{\alpha^2 }{16\pi^2 g} \int_{-\infty}^{\infty} \frac{dk}{2\pi} |k| \big|\mathcal{F}(\text{Si}[x^2+ \frac{y^2}{4t}] )(k)\big|^2 
\end{aligned}
\end{equation}

Appendix \ref{app:infty-plane-alter} gives an alternative route for the calculation and reaches a simpler expression
\begin{equation}
\label{eq:EEE-inf-plane}
\Delta S_n =  \frac{\alpha^2}{8\pi g} \Big\{ 4 \int_{\sqrt{\frac{y^2}{4\pi gt}}}^{\infty} d\lambda \,\frac{1}{\lambda} \big[\text{C}( \lambda ) - \text{S}( \lambda )\big]^2 \Big\}
\end{equation}
where $C$ and $S$ are the Fresnel cos/sin integrals
\begin{equation}
C[z] = \int^z_0 \cos(\frac{\pi}{2} x^2 ) dx \quad S[z] = \int^z_0 \sin(\frac{\pi}{2} x^2 ) dx 
\end{equation}

\subsubsection{Quasi-Particle Interpretation}

\begin{figure}[h]
\begin{minipage}[t]{0.47\linewidth}
\centering
\includegraphics[width=\textwidth]{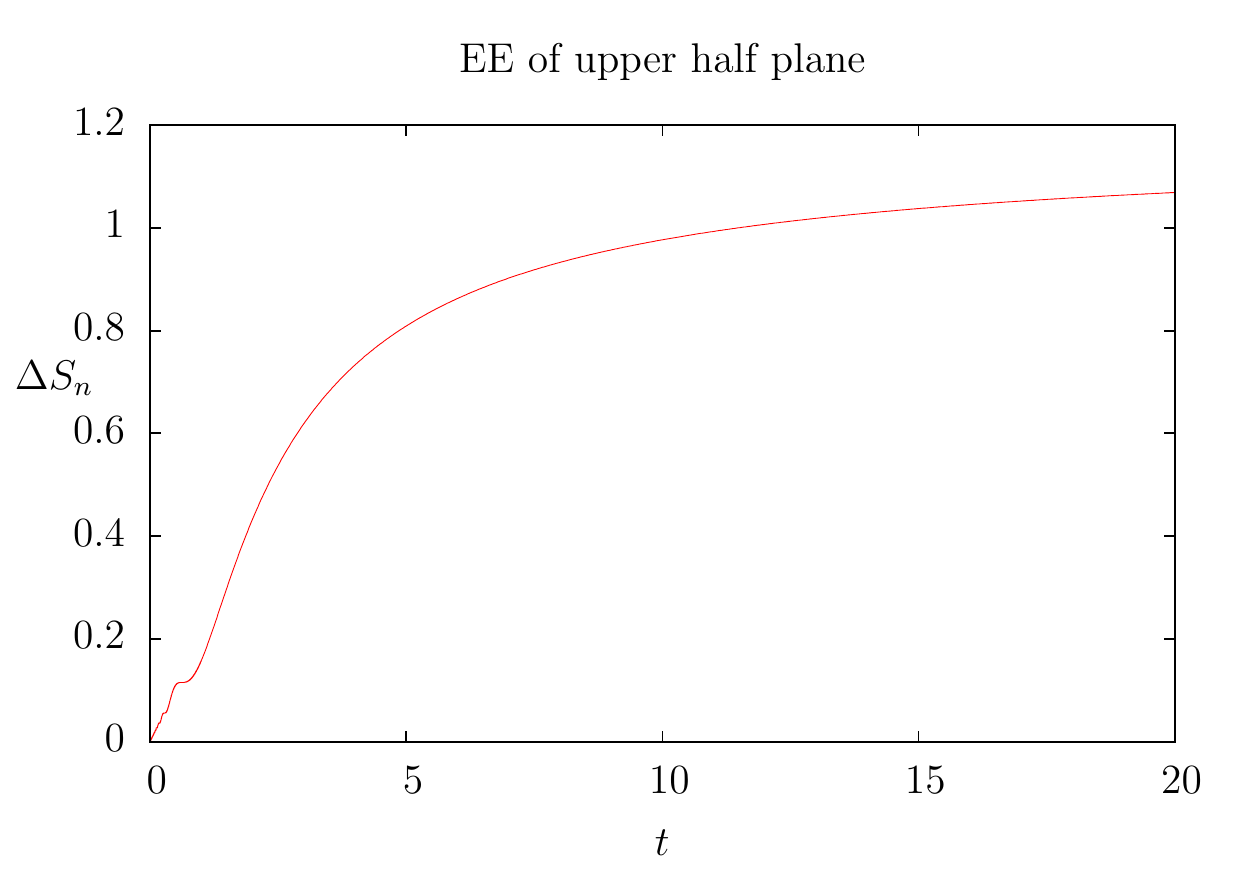}
\caption{\footnotesize Plot of $4 \int_{\sqrt{\frac{1}{t}}}^{\infty} d\lambda \,\frac{1}{\lambda} \big[\text{C}( \lambda ) - \text{S}( \lambda )\big]^2$. EE saturates to constant value in the long time.}
\label{tS.pdf}
\end{minipage}
\hfill
\begin{minipage}[t]{0.47\linewidth}
\centering
\includegraphics[width=\textwidth]{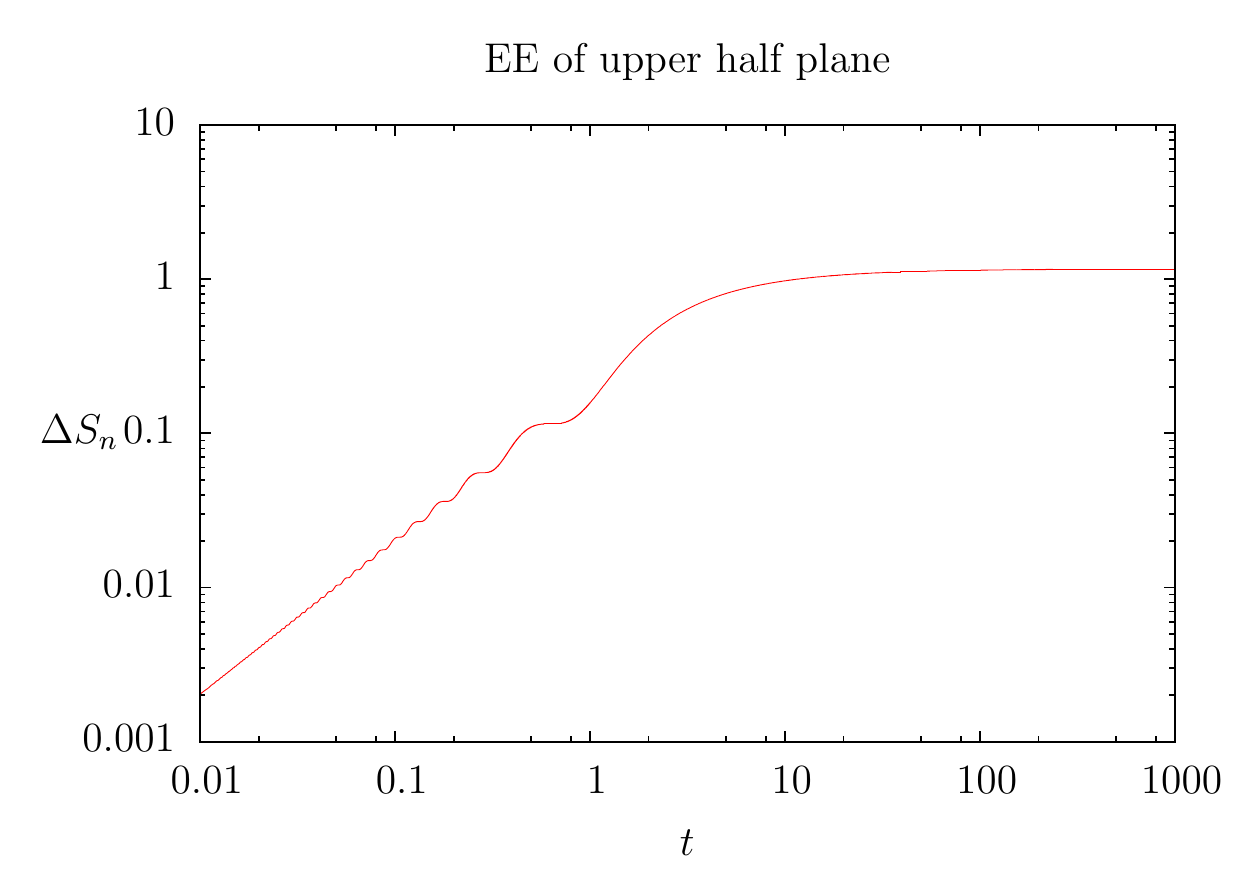}
\caption{This log-log plot exaggerates the plateau in the short time regime. A linear fit in the regime where plateau are invisible gives the slope to be $1$, hence there is a linear increase of excess EE in the short time regime. }
\label{logt_logS.pdf}
\end{minipage}
\end{figure}

We plot the expression of infinite plane excess EE in equation \eqref{eq:EEE-inf-plane} as Fig. \ref{tS.pdf} and \ref{logt_logS.pdf}

Despite some minor modifications, the quasi-particle picture is still able to interpret the growth of EE in this non-relativistic model. 

First of all, the excess EE is a monotonically increasing function of time and eventually saturates to a constant value. The maximal value is proportional to the scaling dimension of the vertex operator, which can be regarded as a dimensionless measure of the strength of the operator. The saturation indicates the exhaustion of quasi-particles, in other words, almost all the downward travelling quasi-particles are in the lower half plane. It is comforting to confirm the fact that the local vertex operator excitation only inject small energies to the system. 

The causality constraint in CFT is superficially violated. The excess EE grows almost immediately after $t = 0$. In fact, the horizon effect is only visible in the regime where $t < \frac{y}{c}$, where $c$ is speed of light (or the equivalent threshold speed in the condensed matter system). While in this non-relativistic theory, the quasi-particle speed is far less than the speed of light, so that we can essentially take the $c \rightarrow \infty$ limit, squeezing the zoom $0 < t < \frac{y}{c}$ to empty. Instead, the typical $z = 2$ diffusive behavior is taking place of the causality constraint. The excess EE grows to $\mathcal{O}(1)$ at the time scale $t \sim y^2$, when the majority of quasi-particle diffuses to the entanglement cut. 

It is interesting to zoom out the small time regime shown in Fig.\ref{logt_logS.pdf}. A linear fit in the log-log plot shows the slope to be $1$ and hence there is a linear increase of excess EE $\Delta S \sim \text{const}\times  \frac{t}{y^2}$ in the short time regime. There are several staircase-like plateau of increasing sizes appear in the growing region, with the last one stacked on top saturating to a limiting value. It is tempting to assume that the quasi-particles disperse: phenomenologically, we see those separated groups of particles arriving sequentially on the entanglement cut. We examine this idea by using a disk probe in the next section. 

\subsection{Disk}

\subsubsection{Excitation in the Center}

If the vertex operator is placed in the center, then the boundary integral kernel function $K( \x_1, \x_2 )$ will only be a function of the angle $\theta$. On the other hand, $\text{Im}f(\x)$, which is the integral of Gaussian, is only a function of the radius. Thus the regularized boundary integral \eqref{eq:cross-Green} is identically zero. 

The vanishing of excess EE in this geometry indicates that the quasi-particle are distributed and travelling with spherical symmetry. Points of excitation away from the center is thus a possible way to probe and decompose the quasi-particle distribution. 

\subsubsection{Using Small Disk as a Probe}

Now we place a disk of radius $r$ centered at the origin, and the excitation at distance $R$ away from the center. The equal time Green function becomes
\begin{equation}
\begin{aligned}
\GDA &= - \frac{1}{2\pi } \ln| z_1 - z_2 | + \frac{1}{2\pi}\ln | z_1 - \frac{R^2}{\bar{z}_2} | \\ 
 & = - \frac{1}{4\pi } \ln\big[r_1 ^2 + r_2 ^2 - 2 r_1 r_2 \cos ( \theta_1 - \theta_2 ) \big] \\
& + \frac{1}{4\pi} \ln\big[r_1 ^2 +  \frac{R^4}{r_2^2} - 2 r_1 \frac{R^2}{r_2} \cos ( \theta_1 - \theta_2 ) \big]
\end{aligned}
\end{equation}
The kernel on the circle is
\begin{equation}
 \partial_{r_1} \partial_{r_2 } \GDA  =  \frac{1}{4\pi R^2  \sin^2 \frac{\theta_1 - \theta_2}{2}}
\end{equation}
Hence
\begin{equation}
\begin{aligned}
 G^{\text{A}}_{\Delta}( \tau_1, \tau_2, \times  )\Big|^{\text{no cut}}_{\text{Dirichlet}}  &= 4 \int_{-\pi}^{\pi} d\theta_1\int_{-\pi}^{\pi} d\theta_2 \frac{\text{Im}f(\x_1)\text{Im}f(\x_2)}{4\pi  \sin^2 \frac{\theta_1 - \theta_2}{2}}\\ 
 &= -4 \int_{-\pi}^{\pi} f( \x ) \partial_{\theta} f(\x)_{\mathcal{H}} d \theta \\
 &= - 4 \frac{1}{2\pi}\sum_{n=-\infty}^{\infty} |n| f_n f_{-n}               
\end{aligned} 
\end{equation}
where $\mathcal{H}$ is the Hilbert transform on circle, and
\begin{equation}
  f( \x )  = \frac{1}{4\pi} \text{Si}\Big[\frac{r^2 + R^2 - 2r R \cos \theta }{4t}\Big]
\end{equation}

Rescaling $R = 1$, we have
\begin{equation}
\label{eq:Sn-disk}
\Delta S_n = \frac{\alpha^2}{4g} 4 \frac{1}{2\pi} \frac{1}{(4\pi)^2 }\sum_{n=-\infty}^{\infty} |n| \text{Si}_n \text{Si}_{-n}
=  \frac{\alpha^2}{8 \pi g }  \frac{1}{4\pi^2 }\sum_{n=-\infty}^{\infty} |n| \text{Si}_n \text{Si}_{-n}
\end{equation}
where $\text{Si}_n$ is the Fourier transform 
\begin{equation}
\text{Si}_n = \int_{-\pi}^{\pi} \text{Si}\Big[\frac{r^2 + 1 - 2r  \cos \theta }{4t}\Big] e^{i n \theta } d\theta
\end{equation}

\subsubsection{Discussion of the Probed Excess EE}

\begin{figure}[h]
\centering
\includegraphics[width=\textwidth]{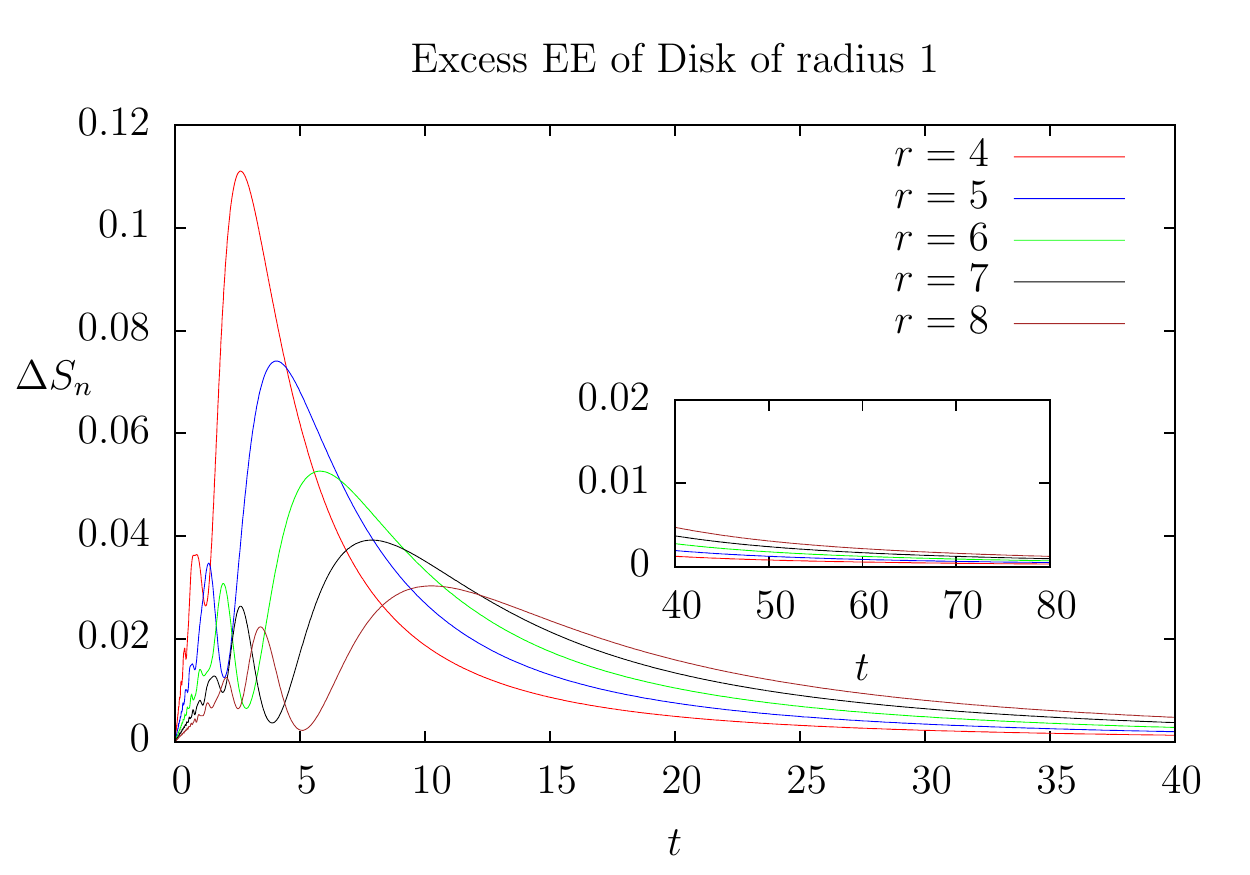}
\caption{Excess EE of disk of radius 1 with different distances to the point of excitation. We can see clearly that the quasi-particle densities are not the same and disperse as time goes on.}
\label{circ_tS.pdf}
\end{figure}

We plot the results in equation \eqref{eq:Sn-disk} in different length scales of distance $r$ to the unit disk probe ($R = 1$). 

In all cases, the excess EE drops down to zero in the asymptotic region when almost all the quasi-particles have passed away. The larger $r$ figure shows larger separations of peaks. There is a largest peak both in height and width that represents the majority of quasi-particles, which should also be responsible for the largest plateau in the upper half plane configuration. Smaller peaks travel faster and arrive earlier to the entanglement cut. This verifies our assumption that the quasi-particles ``wave'' disperse in this diffusion. The pattern is similar to chromatography in chemical species separation which also takes advantage of their different diffusive ``speeds''.  

\section{Summary}
\label{sec:summary}

In this paper we investigate the excess EE created by the vertex operator in quantum Lifshitz model ground state. We develop the replica trick to derive a formula that relate excess EE to the differences of the vertex two point function with Dirichlet boundary and free space. It turns out that excess EE can be completely written in terms of boundary integral on the entanglement cut, and in some sense reflects the fact that the change of EE only happens in the vicinity of the entanglement cut. This is in compliant with the local interaction nature of the original quantum dimer model. 

We pay attention to the upper half plane and disk geometries. We show that the quasi-particle picture can still can interpret the growth of excess EE in this non-relativistic model. In particular, strict causality is replaced by diffusive behavior in the sense that the excess EE will reach a $\mathcal{O}(1)$ scale only when the majority of quasi-particles arrive at the entanglement cut. Zooming out the small time regime in the upper half plane geometry show plateaus in different scales. We ascribe this to the different density of states for quasi-particles of different speeds. By placing a disk probe away from the excitation point, we are able to see the chromatography pattern in the excess EE, which demonstrates possible dispersion and different density of states for particle species. 

Further work can be done to understand the new features discovered in this paper. In the cases we considered, more evidences are needed to account for the plateau structure we find in the short time dynamics of the infinite plane case. We only calculate the single vertex operator excitation and shows that the excess EE is independent of the winding sector, which otherwise plays an important role in the ground state EE. The excess EE of similar operators like $e^{i \phi} + e^{-i\phi}$ will have dependence on the compactification radius, from which we would expect to obtain universal information. There are general questions like the way to obtain Rieb-Robinson bound for the initial development of excess EE for $z = 2$. A holographic picture\cite{nishioka_holographic_2009,nozaki_holographic_2013} in the dual Lifshitz gravity (with Ho\v rava-Lifshitz gravity\cite{horava_quantum_2009} being one of the candidates) would also be helpful in understanding the quench behavior here.

\section{Acknowledgement}

The author would like to acknowledge several researchers who provided valuable help. This project was stimulated by a joint discussion with Masahiro Nozaki and Xueda Wen. They kindly shared their knowledge of the quench problems in CFT. Eduardo Fradkin, Xiao Chen provided critical comments on the quantum Lifshitz model and quench in general. Michael Stone clarified the subtle distributional interpretation in the boundary integral calculation. The author would like to thank them(XW,XC,MS) for reading the manuscript prior to its publication. This work is supported by the National Science Foundation under grant number NSF-DMR-13-06011. 

\appendix
\section{Boundary Reproducing Kernel}
\label{app:bd_delta}

In this appendix, we show that the normal derivative behaves like a boundary reproducing kernel for harmonic functions. 

Let A be a simply connected region and $\GDA$ the associated Green function with Dirichlet boundary condition,
\begin{equation}
- \nabla_{x_1} ^2  \GDA  = \delta^2 ( \x_1 - \x_2 ) \qquad \GDA \Big|_{\x_1 \in \partial A, \x_2 \in A } = 0.
\end{equation}
Let $f( \x )$ to be a function defined on $\partial A$, then 
\begin{equation}
f( \x_2 ) = \lim_{\x_2 \rightarrow \partial A } \int_{\partial A}\Big[ -\partial_{n_1} \GDA \Big] f ( \x_1 ) \, dl_1,
\end{equation}
in other words, $-\partial_{n_1} \GDA$ looks like a delta function on the boundary.

To prove this, we construct a harmonic function $\phi( \x )$ whose boundary value is $f( \x ) $, 
\begin{equation}
-\nabla^2 \phi (\x) = 0 \quad \phi(\x)\Big|_{\partial A} \equiv  f(\x).
\end{equation}
then we express the harmonic function in terms of its boundary value,
\begin{equation}
\begin{aligned}
\phi(\x_2 ) &= \int_A \delta^2( \x_1 - \x_2 )  \phi( \x_1 ) d^2 x_1  =  \int_A (-\nabla_{x_1}^2 )\GDA   \phi( \x_1 ) d^2 x_1 \\
&= - \int_{\partial A} \partial_{n_1} \GDA f ( \x_1 ) \, dl_1 +  \int_A \nabla_{x_1}  \GDA \nabla_{x_1}\phi( \x_1 ) d^2 x_1 \\
&= - \int_{\partial A} \partial_{n_1} \GDA f ( \x_1 ) \, dl_1, 
\end{aligned}
\end{equation}
where we use integration by part twice. 

Taking $\x_2$ approaching the boundary, we obtained the desired identity
\begin{equation}
f( \x_2 ) =  - \lim_{\x_2 \rightarrow \partial A } \int_{\partial A} \partial_{n_1} \GDA f ( \x_1 ) \, dl \\
\end{equation}

We demonstrate this result with the explicit example of upper half plane, whose Green function is
\begin{equation}
\GDA = - \frac{1}{2\pi }\ln |\x_1 -  \x_2| + \frac{1}{2 \pi}\ln | \x_1 - \bar{\x}_2 |. 
\end{equation}
The normal derivative of the green function on the boundary -- the x-axis -- is
\begin{equation}
- \partial_{y_1} \GDA \Big|_{y_1 = 0}  = \frac{1}{\pi} \frac{y_2}{ (x_1 - x_2 )^2 + y_2^2 }. 
\end{equation}
The fact that
\begin{equation}
\int_{-\infty}^{\infty}  dx_2 \frac{1}{\pi} \frac{y_2}{(x_1- x_2 )^2 + y_2^2} = 1
\end{equation}
and
\begin{equation}
  \lim_{y_2 \rightarrow 0 } \frac{1}{\pi} \frac{y_2}{(x_1 - x_2)^2 + y_2^2} = 0 \qquad \text{if } x_1 \ne x_2
\end{equation}
suggest that when $\x_2$ approach the x-axis, 
\begin{equation}
\lim_{y_2 \rightarrow 0 } \frac{1}{\pi} \frac{y_2}{ (x_1 - x_2 )^2 + y_2^2} = \delta( x_1 - x_2 ). 
\end{equation}

\section{Equal Space Green Function for a Simply Connected Region}
\label{app:eq_space_green}

In this appendix, we calculate the derivative of the equal space Green function for a simply connection region A. 

Suppose the equal Green function of Laplacian in region A is denoted as $\GDA$, which satisfies
\begin{equation}
\nabla_{x_1}^2 \GDA = 0 \qquad \GDA\Big|_{\partial A} = 0 .
\end{equation}
The equal space Green function we are seeking for is then 
\begin{equation}
G^{\text{A}}_{\Delta}( \tau_1, \tau_2 ) = \int_{A \times A}  \H 1  G_{\Delta}^{\text{A}}( \x_1, \x_2 )   \H 2  \, d^2 x_1 \, d^2 x_2 
\end{equation}
In the following, we will make use of the heat kernel property
\begin{equation}
\partial_{\tau} H( \x , \tau; \x' ) = \nabla^2 H ( \x, \tau; \x' ) 
\end{equation}
when taking derivatives to the Green function. 

To simplify the notation, we denote
\begin{equation}
\phi(\x_1 ) = \int_A G_{\Delta}^{\text{A}}( \x_1, \x_2 )   \H 2  \, d^2 x_2 
\end{equation}
to be the solution of the Poisson equation in region $A$ with heat kernel as its source 
\begin{equation}
\nabla_{x_1}^2 \phi(\x_1 ) = H( \x, \tau_2; \x_1 ) \qquad \phi(\x_1)\Big|_{\partial A }  = 0 
\end{equation}
then
\begin{equation}
G^{\text{A}}_{\Delta}( \tau_1, \tau_2 ) = \int_{A } \H 1  \phi( \x_1 )  \, d^2 x_1 
\end{equation}

Now taking the $\tau_1$ derivative and integrating by part twice, we have
\begin{equation}
\begin{aligned}
- \partial_{\tau_1 } G^{\text{A}}_{\Delta}( \tau_1, \tau_2 )  &= - \int_A \nabla^2_{x_1} \H 1 \phi( \x_1 ) d^2 x_1  = \int_A \nabla_{x_1} \H 1 \nabla_{x_1}\phi( \x_1 ) d^2 x_1 \\
&=  \int_A  \H 1 ( - \nabla^2_{x_1} ) \phi( \x_1 ) d^2 x_1 + \int_{\partial A }   \H 1 \vec{n}_1\cdot  \nabla_{x_1}\phi( \x_1 ) dl_1\\
&= \int_A  \H 1  H( \x, \tau_2; \x_1 )  d^2 x_1  + \int_{\partial A }   \H 1  \partial_{n_1} \phi( \x_1 ) dl_1 
\end{aligned}
\end{equation}

And then we take symmetrically a $\tau_2$ derivative 
\begin{equation}
\partial_{\tau_2 } \partial_{\tau_1 } G^{\text{A}}_{\Delta}( \tau_1, \tau_2 )  = -\int_A  \H 1  \partial_{\tau_2 } H( \x, \tau_2; \x_1 )  d^2 x_1  -\int_{\partial A }   \H 1  \partial_{n_1}  \partial_{\tau_2 } \phi( \x_1 ) \, dl_1
\end{equation}
and analyze the resulting two terms. 

Upon integration by part, the first term becomes 
\begin{equation}
\begin{aligned}
\int_A  \H 1  (- \nabla_{x_1}^2 ) H( \x, \tau_2; \x_1 ) \, d^2 x_1 &= \int_A \nabla_{x_1} \H 1  \nabla_{x_1} H( \x, \tau_2; \x_1 ) \, d^2 x_1 \\
&- \int_A  \H 1  \partial_{n_1}  H( \x, \tau_2; \x_1 ) \, dl_1
\end{aligned}
\end{equation}
Then we turn to the $\tau_2$ derivative in the second term
\begin{equation}
\begin{aligned}
\partial_{n_1}  \partial_{\tau_2 } \phi( \x_1 )  =& \int_A \partial_{n_1} \GDA (\nabla_{x_2}^2 ) \H 2  \, d^2 x_2 \\
=& \lim_{\x_2 \rightarrow \partial A } \int_A \partial_{n_1} \GDA  \partial_{n_2}  \H 2  \, d^2 x_2 \\
 & - \int_A  \nabla_{x_2} \partial_{n_1} \GDA \nabla_{x_2}  \H 2  \, d^2 x_2 \\
=& -\partial_{n_1}  H( \x, \tau_2; \x_1 )\Big|_{\x_1 \in A} - \int_{\partial A}  \partial_{n_1} \partial_{n_1} \GDA  \H 2  \, dl \\
& + \lim_{\x_1 \in \partial A, \x_2 \rightarrow \partial A } \int_A   \partial_{n_1} (\nabla_{x_2}^2 ) \GDA  \H 2  \, d^2 x_2 \\
=& -\partial_{n_1}  H( \x, \tau_2; \x_1 )\Big|_{\x_1 \in A} - \int_{\partial A}  \partial_{n_1} \partial_{n_1} \GDA  \H 2  \, dl \\
\end{aligned}
\end{equation}

where I have used the theorem in Appendix \ref{app:bd_delta} in going from the second equality to the third. 

Collecting all these results, we have
\begin{equation}
\begin{aligned}
\partial_{\tau_2 } \partial_{\tau_1 } G^{\text{A}}_{\Delta}( \tau_1, \tau_2 ) = &\int_A \nabla_{x_1} \H 1  \nabla_{x_1} H( \x, \tau_2; \x_1 ) \, d^2 x_1 \\
& +\int_{\partial A \times \partial A } \H 1  \partial_{n_1} \partial_{n_2} \GDA  \H 2  \, dl_1 \, dl_2.
\end{aligned}
\end{equation}
The result is symmetric about $\tau_1$ and $\tau_2$. It consists of a contribution purely from the bulk and boundary of region A, where the former is what would be there if the Dirichlet boundary condition were not imposed on the entanglement cut. 

Therefore, compared with the free Green function, we have
\begin{equation}
\label{eq:2nd_der_G}
\partial_{\tau_2 } \partial_{\tau_1 } G^{\text{A}}_{\Delta}( \tau_1, \tau_2 )\Big|^{\text{no cut}}_{\text{Dirichlet}} 
 =  -\int_{\partial A \times \partial A } \H 1  \partial_{n_1} \partial_{n_2} \GDA  \H 2  \, dl_1 \, dl_2.
\end{equation}

\section{Alternative Calculation for the Equal Space Green Function on the Half Plane}
\label{app:infty-plane-alter}

In this section, we consider an alternative calculation of $\Delta S_n$ for upper half plane case. 

\begin{figure}[h]
\centering
\includegraphics[width=0.2\textwidth]{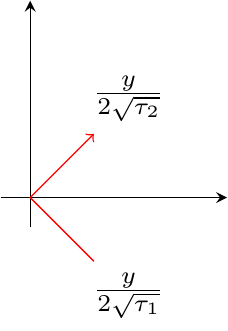}
\caption{We choose the integration path where only the length of the complex number changes.}
\label{fig:int-path}
\end{figure}

According to equation \eqref{eq:eq_space_green_AB}, the general equal space Green function with Dirichlet boundary condition on the entanglement cut for partition A and B is 
\begin{equation}
\begin{aligned}
G_{\Delta}( \tau_1, \tau_2 ) \Big|_{\text{Dirichlet}}
&= \int_{A \times A}d^2 x_1 d^2 x_2 \, H(\vec{x}, \tau_1; \vec{x}_1)  G_\Delta^{A}( \vec{x}_1 , \vec{x}_2 )  H( \vec{x}, \tau_2; \vec{x}_2 ) \\
& + \int_{B \times B}d^2 x_1 d^2 x_2 \, H(\vec{x}, \tau_1; \vec{x}_1)  G_\Delta^{B}( \vec{x}_1 , \vec{x}_2 )  H( \vec{x}, \tau_2; \vec{x}_2 ) \\
\end{aligned}
\end{equation}
The difference of the presence and absence of the cut will only come from the boundaries of $A$ and $B$, which is the x-axis. By applying the general formula \eqref{eq:2nd_der_G} for both region A and B, we have
\begin{equation}
\partial_{\tau_2} \partial_{\tau_1} G_\Delta( \tau_1, \tau_2 ) \Big|^{A \cup B }_{\text{Dirichlet}} = - 2\int_{-\infty}^{\infty} d x_2 \int_{-\infty}^{\infty} d x_1 \, \H 1  \partial_{y_1} \partial_{y_2} \GD  \H 2 \Big|_{y_1 = 0,y_2\rightarrow  0 }
\end{equation}
We interpret(or regulate) this boundary integral as in Appendix \ref{app:distri-bd-int}
\begin{equation}
\label{eq:tt_deri_G_diff}
\partial_{\tau_2} \partial_{\tau_1} G_\Delta( \tau_1, \tau_2 ) \Big|^{A\cup B}_{\text{Dirichlet}} = \frac{1}{\pi}\int dx_1 dx_2  \frac{\big[\H 1  - H( \x, \tau_1; \x_2 ) \big]\big[ H( \x, \tau_2; \x_1 ) - \H 2 ]}{ (x_1 - x_2 )^2 } 
\end{equation}
The calculation is a little involved, so for clarity we provide the details in Appendix \ref{app:sch_para_calc}, and the result is
\begin{equation}
\partial_{\tau_2} \partial_{\tau_1} G_\Delta( \tau_1, \tau_2 ) \Big|^{A\cup B}_{\text{Dirichlet}}  = \frac{1}{4\pi^2} \frac{1}{\tau_1 + \tau_2 } \frac{1}{\sqrt{\tau_1}}\exp(- \frac{y^2}{4\tau_1}) \frac{1}{\sqrt{\tau_2}}\exp(- \frac{y^2}{4\tau_2})
\end{equation}
The Green function can be obtained by integrating of $\tau_1$ and $\tau_2$,
\begin{equation}
\begin{aligned}
G_\Delta( \tau_1, \tau_2 ) \Big|^{A\cup B}_{\text{Dirichlet}}  &=\frac{1}{4\pi^2} \int d\tau_1 d\tau_2 \frac{1}{\tau_1 + \tau_2 } \frac{1}{\sqrt{\tau_1}}\exp(- \frac{y^2}{4\tau_1}) \frac{1}{\sqrt{\tau_2}}\exp(- \frac{y^2}{4\tau_2}) \\
&=\frac{1}{\pi^2} \int_0^{\sqrt{\tau_1} }du  \int_0^{\sqrt{\tau_2}} dv \exp( - \frac{y^2}{4u^2} )\frac{1}{u^2 + v^2 }\exp( - \frac{y^2}{4v^2} ) \\
&=\frac{1}{\pi^2} \int_0^{2\sqrt{\tau_1} }du  \int_0^{2\sqrt{\tau_2}} dv \exp( - \frac{y^2}{u^2} )\frac{1}{u^2 + v^2 }\exp( - \frac{y^2}{v^2} )  \\
&=\frac{1}{\pi^2} \int_{\frac{1}{2\sqrt{\tau_1}}}^{\infty}du  \int_{\frac{1}{2\sqrt{\tau_2}}}^{\infty} dv \exp( -u^2 y^2 )\frac{1}{u^2 + v^2 }\exp( - v^2 y^2  )\\
&=\frac{1}{\pi^2} \int_{\frac{y}{2\sqrt{\tau_1}}}^{\infty}du  \int_{\frac{y}{2\sqrt{\tau_2}}}^{\infty} dv \exp( -u^2 )\frac{1}{u^2 + v^2 }\exp( - v^2 )
\end{aligned}
\end{equation}
The excess EE is proportional to the cross Green function, which is
\begin{equation}
\begin{aligned}
G_{\Delta}(\tau_1, \tau_2, \times) = &\Big\{G_\Delta( \tau_1, \tau_1 ) - G_\Delta( \tau_1, \tau_2 ) - G_\Delta( \tau_2, \tau_1 ) + G_\Delta( \tau_2, \tau_2 ) \Big\}\Big|^{A\cup B}_{\text{Dirichlet}} \\
=&\frac{1}{\pi^2} \int_{\frac{y}{2\sqrt{\tau_1}}}^{\frac{y}{2\sqrt{\tau_2}}}du \int_{\frac{y}{2\sqrt{\tau_1}}}^{\frac{y}{2\sqrt{\tau_2}}} dv \exp( -u^2)\frac{1}{u^2 + v^2 }\exp( - v^2  )\\
=& \frac{1}{\pi^2} \int_0^{\infty} d\lambda \, \Big\{ \int_{\frac{y}{2\sqrt{\tau_1}}}^{\frac{y}{2\sqrt{\tau_2}}}du  \exp( -(1+ \lambda)u^2) \Big\}^2 \\
=& \frac{1}{\pi^2} \int_0^{\infty} d\lambda \,\frac{1}{1+\lambda} \Big\{ \int_{\frac{\sqrt{1+ \lambda} y}{2\sqrt{\tau_1}}}^{\frac{\sqrt{1+ \lambda} y}{2\sqrt{\tau_2}}}du  \exp( -u^2) \Big\}^2 
\end{aligned}
\end{equation}
We calculate the contour integral in the curly braces as in Figure \ref{fig:int-path}. The path is $\frac{\sqrt{1+ \lambda} y}{2\sqrt{\tau_1}} \rightarrow 0 \rightarrow \frac{\sqrt{1+ \lambda} y}{2\sqrt{\tau_2}}$, whose two segments only changes the length of the complex number. 

\begin{equation}
\begin{aligned}
\int_{\frac{\sqrt{1+ \lambda} y}{2\sqrt{\tau_1}}}^{\frac{\sqrt{1+ \lambda} y}{2\sqrt{\tau_2}}} du  \exp( -u^2)  &= \int_{r_0}^0  du\, \sqrt{-i} \exp( i u^2 ) + \int_0^{r_0} du \, \sqrt{i} \exp( -i u^2 ) \\
&= \sqrt{2}i \big[ \int_0^{r_0} du \, \cos u^2 - \int_0^{r_0} \sin u^2 du \big]\\
&= \sqrt{2}i  \sqrt{\frac{\pi}{2}} \big[\text{C}( \sqrt{\frac{2}{\pi}} r_0 ) - \text{S}( \sqrt{\frac{2}{\pi}} r_0 )\big]
\end{aligned}
\end{equation}
where $C$ and $S$ are the Fresnel cos/sin integrals
\begin{equation}
C[z] = \int^z_0 \cos(\frac{\pi}{2} x^2 ) dx \quad S[z] = \int^z_0 \sin(\frac{\pi}{2} x^2 ) dx 
\end{equation}
and the constant $r_0 =\frac{\sqrt{1+ \lambda} y}{2\sqrt{2gt}}$. 

This gives the cross Green function 
\begin{equation}
\begin{aligned}
G_{\Delta}(\tau_1, \tau_2, \times) &= -\frac{1}{\pi}   \int_0^{\infty} d\lambda \,\frac{1}{1+\lambda} \bigg[\text{C}( \sqrt{\frac{2}{\pi}} r_0 ) - \text{S}( \sqrt{\frac{2}{\pi}} r_0 )\bigg]^2 \\
&= - \frac{1}{\pi}\int_1^{\infty} d\lambda \,\frac{1}{\lambda} \bigg[\text{C}( \sqrt{\frac{\lambda}{4\pi g t}} y ) - \text{S}( \sqrt{\frac{\lambda}{4\pi g t}} y  )\bigg]^2 \\
&= - \frac{1}{\pi} \int_{\frac{y^2}{4\pi g t}}^{\infty} d\lambda \,\frac{1}{\lambda} \big[\text{C}( \sqrt{\lambda} ) - \text{S}( \sqrt{\lambda}   )\big]^2 \\
&= - \frac{2}{\pi} \int_{\sqrt{\frac{y^2}{4\pi g t}}}^{\infty} d\lambda \,\frac{1}{\lambda} \big[\text{C}( \lambda ) - \text{S}( \lambda )\big]^2 \\
\end{aligned}
\end{equation}
and consequently the excess EE
\begin{equation}
\label{eq:UHP-t-inf}
\begin{aligned}
\Delta S_n 
&= - \frac{\alpha^2}{4g} G_{\Delta}(\tau_1, \tau_2, \times)
&= \frac{\alpha^2}{8\pi g} \Big[ 4 \int_{\sqrt{\frac{y^2}{2\pi t}}}^{\infty} d\lambda \,\frac{1}{\lambda} \big[\text{C}( \lambda ) - \text{S}( \lambda )\big]^2 \Big]\\
\end{aligned}
\end{equation}

\section{Schwinger Parameter Calculation}
\label{app:sch_para_calc}

This section is devoted to the analytic calculation of equation \eqref{eq:tt_deri_G_diff}. 

After doing the proper regularization, the integral in equation \eqref{eq:tt_deri_G_diff} is convergent, so it is safe to add a small positive constant $\epsilon$ in
\begin{equation}
\partial_{\tau_2} \partial_{\tau_1} G_\Delta( \tau_1, \tau_2 ) \Big|^{A\cup B}_{\text{Dirichlet}} = \frac{1}{\pi}\lim_{\epsilon\rightarrow 0^+} \int dx_1 dx_2  \frac{\big[\H 1  - H( \x, \tau_1; \x_2 ) \big]\big[ H( \x, \tau_2; \x_1 ) - \H 2 ]}{(x_1 - x_2 )^2  + \epsilon} 
\end{equation}
to compute each term. The divergent pieces in the limit $\epsilon \rightarrow 0^+$ will automatically cancel in the end. 

The four integrals are independent of $x$,
\begin{equation}
\begin{aligned}
  \frac{1}{\pi}\int dx_1 dx_2  \frac{H( \x, \tau_1; \x_i) H( \x, \tau_2; \x_j ) }{(x_1 - x_2 )^2  + \epsilon}  &= \frac{1}{(4\pi)^2 \tau_1 \tau_2 }\exp\big[- (\frac{1}{4\tau_1} + \frac{1}{4\tau_2})y^2\big]I_{ij}
\end{aligned}
\end{equation}
where
\begin{equation}
I_{ij} = \frac{1}{\pi}\int dx_1 dx_2 \frac{\exp(-\frac{1}{4\tau_1}x_i^2 - \frac{1}{4\tau_2} x_j^2 )}{(x_1- x_2 )^2 + \epsilon}
\end{equation}
such that
\begin{equation}
\begin{aligned}
\partial_{\tau_2} \partial_{\tau_1} G_\Delta( \tau_1, \tau_2 ) \Big|^{A\cup B}_{\text{Dirichlet}} &= \frac{1}{(4\pi)^2 \tau_1 \tau_2 }\exp\big[- (\frac{1}{4\tau_1} + \frac{1}{4\tau_2})y^2\big] \lim_{\epsilon\rightarrow 0^+} I_{11} - I_{12} - I_{21} + I_{22}  \\
&= \frac{1}{(4\pi)^2 \tau_1 \tau_2 }\exp\big[- (\frac{\tau_1 + \tau_2}{4\tau_1\tau_2})y^2\big] \lim_{\epsilon\rightarrow 0^+} 2( I_{11} - I_{12} )
\end{aligned}
\end{equation}
The integral $I_{11}$ can be done directly
\begin{equation}
\begin{aligned}
I_{11} &= \int dx_1 \exp(-(\frac{1}{4\tau_1}+ \frac{1}{4\tau_2}) x_1^2 ) \frac{1}{\sqrt{\epsilon} } \frac{1}{\pi}\int dx_2 \frac{\sqrt{\epsilon}}{ (x_1 - x_2)^2 + \epsilon} \\
&= \sqrt{\frac{\pi}{\epsilon}}\sqrt{\frac{4\tau_1 \tau_2}{\tau_1 + \tau_2}}
\end{aligned}
\end{equation}
$I_{12}$ can be done using Schwinger parameter
\begin{equation}
\begin{aligned}
I_{12} &= \int_0^{\infty}  d\lambda \, \frac{1}{\pi} \int dx_1 dx_2 \exp\big[-\frac{1}{4\tau_1}x_1^2 - \frac{1}{4\tau_2} x_2^2 - \lambda (x_1 - x_2)^2 - \lambda \epsilon \big]\\
&= \int_0^{\infty}  d\lambda \, \frac{1}{\sqrt{\frac{1}{4\tau_1} + \lambda }}\frac{\sqrt{\frac{1}{4\tau_1} + \lambda }}{\sqrt{(\frac{1}{4\tau_1}+ \frac{1}{4 \tau_2})\lambda + \frac{1}{16 \tau_1 \tau_2 } }} e^{-\lambda \epsilon} \\
&= \sqrt{\frac{4\tau_1 \tau_2}{\tau_1 + \tau_2}}\int_0^{\infty} d \lambda \frac{1}{\sqrt{ \lambda + \frac{1}{4 (\tau_1 + \tau_2 )}}} e^{- \lambda \epsilon } 
\end{aligned}
\end{equation}
We are able to do the following indefinite integral
\begin{equation}
\int d\lambda \frac{1}{\sqrt{\lambda + \lambda_0 }}e^{- \epsilon\lambda }  = -\sqrt{\frac{\pi}{\epsilon}} e^{\lambda_0 \epsilon} \text{erfc}( \sqrt{ \epsilon( \lambda + \lambda_0) }).
\end{equation}
At $\lambda \rightarrow \infty$, $\text{erfc} \rightarrow 0 $, so 
\begin{equation}
I_{12} =  \sqrt{\frac{4\tau_1 \tau_2}{\tau_1 + \tau_2}} \sqrt{\frac{\pi}{\epsilon}} e^{\lambda_0 \epsilon} \text{erfc}( \sqrt{ \epsilon \lambda_0 })
\end{equation}
where 
\begin{equation}
\lambda_0 = \frac{1}{4( \tau_1 + \tau_2 ) }
\end{equation}
When taking the $\epsilon \rightarrow 0 $ limit, $\text{erfc}(x) \sim 1 - \frac{2x}{\sqrt{\pi}}$
\begin{equation}
I_{12} = \sqrt{\frac{4\tau_1 \tau_2}{\tau_1 + \tau_2}} \sqrt{\frac{\pi}{\epsilon}} (1 - \frac{2}{\sqrt{\pi}} \sqrt{ \epsilon \lambda_0} )
\end{equation}
therefore
\begin{equation}
\lim_{\epsilon\rightarrow 0^+} I_{11} - I_{12} =  \sqrt{\frac{4\tau_1 \tau_2}{\tau_1 + \tau_2}}  2 \sqrt{\lambda_0}  =  \frac{\sqrt{4 \tau_1 \tau_2}}{\tau_1 + \tau_2} 
\end{equation}
and the Green function becomes
\begin{equation}
\partial_{\tau_2} \partial_{\tau_1} G_\Delta( \tau_1, \tau_2 ) \Big|^{A\cup B}_{\text{Dirichlet}}  = \frac{1}{4\pi^2} \frac{1}{\tau_1 + \tau_2 } \frac{1}{\sqrt{\tau_1}}\exp(- \frac{y^2}{4\tau_1}) \frac{1}{\sqrt{\tau_2}}\exp(- \frac{y^2}{4\tau_2})
\end{equation}

\section{Distributional Boundary Integral}
\label{app:distri-bd-int}
In this section, we focus on the correct distributional interpretation of integral of the type (see exercise 6.8 of the book \cite{stone_mathematics_2009})
\begin{equation}
I = \lim_{y_2 \rightarrow 0 } \int_{-\infty}^{\infty} d x_1 \int_{-\infty}^{\infty} d x_2 \, f(x_1 )   \partial_{y_2}\partial_{y_1}   \GD \Big|_{y_1 = 0} g(x_2 ) 
\end{equation}
where the kernel of the integral is
\begin{equation}
\begin{aligned}
\lim_{y_2\rightarrow 0} \partial_{y_2} \partial_{y_1} \GD \Big|_{y_1 = 0 }  &= \frac{1}{\pi}\lim_{y_2\rightarrow 0}\partial_{y_2}  \frac{y_2 }{(x_1- x_2 )^2 + y_2^2} \\
& = \frac{1}{\pi}\lim_{y_2\rightarrow 0} \Big\{ \frac{1}{(x_1- x_2 )^2 + y_2^2} - \frac{2y_2^2 }{[(x_1- x_2 )^2 + y_2^2]^2 } \Big\} 
\end{aligned}
\end{equation}

It is well approximated by $\frac{1}{(x_1 - x_2)^2}$ when $|x_1 - x_2 | \ge y_2$, but is singular in the other limit. The fact that
\begin{equation}
  \int_{-\infty}^{\infty} dx_1\, \frac{1}{\pi} \frac{y_2}{(x_1 - x_2 )^2 + y_2^2}  = 1 
\end{equation}
is independent of $y_2$ suggests
\begin{equation}
\int_{-\infty}^{\infty} \partial_{y_2}  \frac{1}{\pi} \frac{y_2}{(x_1 - x_2 )^2 + y_2^2}  = 0
\end{equation}
Hence there is a highly localized distribution at $x_1 = x_2$ to compensate the (positive) divergent integral of $P \int_{-\infty}^{\infty} dx_1 \frac{1}{(x_1 - x_2 ) ^2}$. 

We notice that
\begin{equation}
\partial_{y_2}\partial_{y_1}   \GD \Big|_{y_1 = 0} = \frac{1}{\pi} \frac{(x_1-x_2)^2 - y_2^2 }{[(x_1- x_2 )^2 + y_2^2]^2 } = \partial_{x_2} \frac{1}{\pi}   \frac{x_1-x_2}{(x_1- x_2 )^2 + y_2^2}
\end{equation}
and hence
\begin{equation}
I = \frac{1}{\pi} \lim_{\epsilon \rightarrow 0^+ } \int_{-\infty}^{\infty} d x_1 \int_{-\infty}^{\infty} d x_2 \, f(x_1)   g(x_2 ) \partial_{x_2}    \frac{x_1-x_2}{(x_1- x_2 )^2 + \epsilon}  
\end{equation}
The rewriting does not remove the singularity; what we do is to integrate the kernel in region $|x_1 - x_2| > \Lambda$, where the it can be approximated by $\frac{1}{(x_1 - x_2 )^2}$ and $| x_1 - x_2| < \Lambda$ where $g(x_2 ) \sim g(x_1)$:
\begin{equation}
\begin{aligned}
\pi I(\epsilon)  &= \int_{|x_2 - x_2 | > \Lambda}  d x_1d x_2 \frac{f(x_1)   g(x_2 )}{(x_1 - x_2 )^2+ \epsilon }  + \int d x_1 \, f(x_1)   g(x_1 ) \int_{|x_2 - x_1 | \le \Lambda} dx_2 \, \partial_{x_2}    \frac{x_1-x_2}{(x_1- x_2 )^2 + \epsilon}  \\
 &= \int_{|x_2 - x_2 | > \Lambda}  d x_1d x_2 \frac{f(x_1)   g(x_2 )}{(x_1 - x_2 )^2+ \epsilon }  - \int d x_1 \, f(x_1)   g(x_1 ) \int_{|x_2 - x_1 | > \Lambda} dx_2 \, \partial_{x_2}    \frac{x_1-x_2}{(x_1- x_2 )^2 + \epsilon}  \\
 &= \int_{|x_2 - x_2 | > \Lambda}  d x_1d x_2 \frac{f(x_1)   g(x_2 )}{(x_1 - x_2 )^2+ \epsilon }  - \int_{|x_2 - x_1 | > \Lambda} d x_1 dx_2 \frac{f(x_1)   g(x_1 )}{(x_1 - x_2 )^2+ \epsilon } \\
\end{aligned}
\end{equation}
Take $\epsilon \rightarrow 0$ limit and symmetrimize functions
\begin{equation}
I =  -\frac{1}{2} \int dx_1 dx_2  \frac{\big[f(x_1) - f(x_2 )\big]\big[ g(x_1)- g(x_2 )]}{\pi (x_1 - x_2 )^2 } 
\end{equation}
notice that the singularity at $x_1 = x_2$ has been regulated away.

\bibliographystyle{unsrt}
\bibliography{Lifshitz_quench}

\end{document}